\documentclass[11pt,a4]{article}
\usepackage{amsmath}
\usepackage{amscd}
\usepackage{array}
\usepackage{latexsym,amssymb}
\DeclareFontFamily{U}{rsf}{}
\DeclareFontShape{U}{rsf}{m}{n}{
  <5> <6> rsfs5 <7> <8> <9> rsfs7 <10-> rsfs10}{}
\DeclareMathAlphabet\Scr{U}{rsf}{m}{n}

\newcommand{\nn}{\nonumber}

\def\to{\rightarrow}

\newcommand{\be}{\begin{equation}}
\newcommand{\ee}{\end{equation}}
\newcommand{\bea}{\begin{eqnarray}}
\newcommand{\eea}{\end{eqnarray}}

\newcommand{\ba}{\begin{eqnarray}}
\newcommand{\ea}{\end{eqnarray}}

\def\e{\epsilon}

\def\cG{\mathcal{G}}

\def\cL{\mathcal{L}}
\def\cM{\mathcal{M}}

\def\cR{\mathcal{R}}

\def\notin{\hbox{{$\in$}\kern-.51em\hbox{/}}}

\def\inbar{\vrule height1.5ex width.4pt depth0pt}
\def\IB{\relax{\rm I\kern-.18em B}}
\def\IC{\relax\,\hbox{$\inbar\kern-.3em{\rm C}$}}
\def\ID{\relax{\rm I\kern-.18em D}}
\def\IE{\relax{\rm I\kern-.18em E}}
\def\IF{\relax{\rm I\kern-.18em F}}
\def\IG{\relax\,\hbox{$\inbar\kern-.3em{\rm G}$}}
\def\IH{\relax{\rm I\kern-.18em H}}
\def\II{\relax{\rm I\kern-.17em I}}
\def\IK{\relax{\rm I\kern-.18em K}}
\def\IL{\relax{\rm I\kern-.18em L}}
\def\IN{\relax{\rm I\kern-.18em N}}
\def\IP{\relax{\rm I\kern-.18em P}}
\def\IQ{\relax\,\hbox{$\inbar\kern-.3em{\rm Q}$}}
\def\IR{\relax{\rm I\kern-.18em R}}
\def\IU{\relax\,\hbox{$\inbar\kern-.3em{\rm U}$}}
\def\ZZ{\relax\ifmmode\mathchoice{\hbox{\cmss
Z\kern-.4em Z}}{\hbox{\cmss Z\kern-.4em Z}}{\lower.9pt\hbox{\cmsss Z\kern-.4em Z}} {\lower1.2pt\hbox{\cmsss
Z\kern-.4em Z}}\else{\cmss Z\kern-.4em Z}\fi}
\def\IGam{\relax{{\rm I}\kern-.18em \Gamma}}

\def\bfnull{\relax{\rm O \kern-.635em 0}}

\def\square{{\,\lower0.9pt\vbox{\hrule
\hbox{\vrule height 0.2 cm \hskip 0.2 cm \vrule height 0.2 cm}\hrule}\,}}

\def\twomat#1#2#3#4{\left(\begin{array}{cc} \end{array} \right)}

\begin{document}

\numberwithin{equation}{section}

\begin{center}
{\bf\LARGE Extremal Limits of Rotating Black Holes} \\
\vskip 2 cm
{\bf \large Laura Andrianopoli, Riccardo D'Auria, Antonio Gallerati \\and Mario Trigiante}
\vskip 8mm
 \end{center}
\noindent {\small{\it DISAT, Politecnico di Torino, Corso Duca
    degli Abruzzi 24, I-10129 Turin, Italy and Istituto Nazionale di
    Fisica Nucleare (INFN) Sezione di Torino, Italy}

\vskip 2 cm
\begin{center}
{\small {\bf Abstract}}
\end{center}
We consider non-extremal, stationary,  axion-dilaton solutions to ungauged symmetric supergravity models, obtained by Harrison transformations of the non-extremal Kerr solution. We define a general algebraic  procedure, which can be viewed as an  In\"on\"u--Wigner contraction of the Noether-charge matrix associated with the effective $D=3$ sigma-model description of the solution, yielding, through different singular limits, the known BPS and non-BPS extremal black holes (which include the under-rotating non-BPS one). The non-extremal black hole can thus be thought of as ``interpolating'' among these limit-solutions. The algebraic procedure that we define generalizes
the known Rasheed-Larsen limit which yielded, in the Kaluza-Klein theory, the first instance of under-rotating extremal solution.  As an example of our general result, we discuss in detail the non-extremal solution in the $T^3$-model, with either $q_0,\,p^1$ or $p^0,\,q_1$ charges switched on, and its singular limits. Such solutions, computed in $D=3$ through the solution-generating technique, is completely described in terms of $D=4$ fields, which include the fully integrated vector fields.

\vskip 1 cm
\vfill
\noindent {\small{\it
    E-mail:  \\
    {\tt
      laura.andrianopoli@polito.it}; \\
      {\tt riccardo.dauria@polito.it};\\
      {\tt antonio.gallerati@polito.it}; \\
      {\tt mario.trigiante@polito.it}
      }
   \eject

%%%%%%%%%%%%%%%%%%%%%%%%%%%%%%%%%%%%%%%%%%%%%%%%%%%%%
%%%%%%%%%%%%%%%%%%%%%%%%%%%%%%%%%%%%%%%%%%%%%%%%%%%%%
%%%%%%%%%%%%%%%%%%%%%%%%%%%%%%%%%%%%%%%%%%%%%%%%%%%%%

%\begin{center}
%\today
%\end{center}

%%%%%%%%%%%%%%%%%%%%%%%%%%%%%%%%%%%%%%%%%%%%%%%%%%%%
\section{Introduction}
The study of stationary black holes in superstring/supergravity theories is a branch of research which has witnessed important progresses in the last two decades or so \cite{reviews}. Initially, special attention was devoted to BPS and in general extremal solutions by virtue of their universal near-horizon behavior, due to the \emph{attractor phenomenon} \cite{attractors}. Multicenter extremal solutions in $D=4$  have also been extensively studied in recent years \cite{Goldstein:2008fq,Bena:2009ev,Bossard:2011kz,Ferrara:2012qm}.\par
Our knowledge of non-extremal, stationary  solutions is more restricted,
due to the less constrained form of the  space-time metric.
 The known examples are typically obtained through the so called \emph{solution-generating techniques} \cite{Cvetic:1995kv}.
 The idea underlying this approach is that stationary solutions to $D=4$ supergravity are also solutions to an Euclidean theory in three dimensions, formally obtained by compactifying the $D=4$ parent model  along the time-direction \cite{Breitenlohner:1987dg} and dualizing all the vector fields into scalars. The resulting $D=3$ theory is a sigma-model coupled to gravity and has the desirable feature of having a larger global symmetry group than the original four-dimensional model. The extra symmetries can be used to generate new  four-dimensional solutions from known ones. These symmetries, for instance, include the Harrison transformations which can generate electric and magnetic charges when acting on a neutral solution  like the Schwarzschild or the Kerr black hole. The  relevant physical properties of stationary black holes in four dimensions are thus conveniently described by the  orbits of such solutions with respect to the action of the $D=3$ global symmetry group $G_{(3)}$. \par
 It is a commonly accepted statement in the black-hole literature that extremal solutions in supergravity can be obtained as limits of non-extremal ones. In the extremal limit a certain \emph{extremality parameter}, related to the Hawking temperature of the black hole, is sent to zero. Limits yielding BPS (rotating and non-rotating) solutions were first  defined in  \cite{Cvetic:1996kv}. An other non-trivial example of extremal limit was defined in  \cite{Rasheed:1995zv} (which we shall refer to as the Rasheed-Larsen limit), and allowed to find the first instance of extremal \emph{under-rotating} (i.e. without ergosphere) solutions from a given non-extremal one in the $D=4$ theory obtained, through Kaluza--Klein reduction, from pure gravity in $D=5$. In \cite{Bena:2009ev} it was shown that the extremal  solution  to the Kaluza-Klein theory was duality-related to the class of solutions studied in the same work.
 In \cite{Andrianopoli:2012ee} we generalized the Rasheed-Larsen limit to a non extremal stationary black hole in the $T^3$ model, obtaining the non-BPS under-rotating solution through a singular Harrison transformation on a non-extremal Kerr one. \par
 The aim of the present note is to generalize this limit prescription to generic symmetric, extended supergravities in order to obtain the most general, single center, extremal solution modulo the action of the global symmetry group $G_{(3)}$ (\emph{seed solution} with respect to the action of $G_{(3)}$). To this end we construct a non-extremal rotating, asymptotically-flat black hole by acting by means of suitable  Harrison transformations on the non-extremal Kerr solution. This transformation should switch on the minimal set of charges which allows to obtain, in the appropriate limit, the representatives of the $G_{(3)}$-orbits of all the  extremal BPS and non-BPS $D=4$ solutions.\par
 We shall restrict ourselves to axisymmetric, stationary, asymptotically flat solutions. A valuable tool in order to study the effect of global symmetry transformations on a solution is the Noether-charge matrix $Q$ which encodes, in the $D=3$ description,
the ADM-mass, electric-magnetic charges, NUT charge and scalar charges at infinity. Another valuable tool is the constant matrix $Q_\psi$, first introduced in \cite{Andrianopoli:2012ee}, which is associated with the rotation of the solution and is defined as:
\begin{equation}
Q_\psi =-\frac{3}{4\pi}\,\int_{S_2^\infty}\psi_{[i}\,J_{
j]}\,dx^i\wedge dx^j\,,
\end{equation}
$J_i$, $i=1,2,3$  being the conserved $D=3$ Noether-current and $\psi=\partial_\varphi$ the angular Killing vector. Both $Q$ and $Q_\psi$ belong to the Lie algebra of $G_{(3)}$ and thus transform covariantly under its adjoint action. The angular momentum $M_\varphi$ coincides with the  component of $Q_\psi$  along a  specific generator.
 In \cite{Andrianopoli:2012ee} the regularity condition for (non)-extremal solutions was written in terms of $Q$ and $Q_\psi$ in a $G_{(3)}$-invariant fashion:
 \begin{equation}
m^2\ge \alpha^2\,\,\Leftrightarrow \,\,\,\,{\rm Tr}[Q^2]\ge \frac{2}{k}\,\frac{{\rm Tr}[Q_\psi^2]}{{\rm Tr}[Q^2]}\,,\label{regcond0}
\end{equation}
 Regular non-extremal solutions are characterized by semisimple $Q,\,Q_\psi$, while these matrices, for extremal solutions, are nilpotent \cite{pioline,Gaiotto:2007ag}.
  There have been considerable efforts toward the classification of extremal solutions in supergravity by means of nilpotent orbits \cite{nilorbits} of $G_{(3)}$ \cite{Bergshoeff:2008be,Bossard:2009at,Kim:2010bf,marioetal,Bossard:2011kz,Fre:2011uy,Fre:2011ns,Chemissany:2012nb,Fre:2012im}.
As we shall show, the limit prescriptions that we define, and which yield regular extremal solutions, are such that both $Q$
 and $Q_\psi$ become nilpotent, but $Q_\psi$ has a lower degree of nilpotenty than $Q$, so that both sides of (\ref{regcond0}) vanish separately.\par
 By generalizing the original Rasheed-Larsen limit, our prescription amounts to sending the Harrison parameters either to $+\infty$ or to $-\infty$, and, at the same time, sending the mass, as well as the ratio of the angular momentum to the mass of the original Kerr solution, to zero.
 The effect of this operation is an In\"on\"u-Wigner contraction of $Q$ and $Q_\psi$ which turns them from semisimple to nilpotent.  Depending on which of the Harrison parameters are sent to plus and which to minus infinity, the resulting $Q$ matrices  provide representatives of the relevant $G_{(3)}$-orbits of all the regular extremal solutions. The original non-extremal rotating solution can thus be viewed as  ``interpolating'' among the different extremal BPS and non-BPS limits and, as long as we consider seed solutions with respect to the action of $G_{(3)}$, it also has a universal character. Indeed representatives of the $G_{(3)}$-orbits of regular extremal solutions in the maximal, half-maximal, as well as $\mathcal{N}=2$  models with rank-3 symmetric special K\"ahler manifold, can be obtained as limits of a single non-extremal rotating solution to the $STU$ model. These non-extremal solutions were also considered in \cite{Cvetic:1995kv,Bouchareb:2007ax,Bertini:2011ga,Cvetic:2011dn,Virmani:2012kw}, although for a different purpose.\footnote{See also, in the same context, Ref. \cite{Chakraborty:2012fx}.}\par
 The paper is organized as follows. In Sect. 2 we describe the general mathematical  setting and review the solution-generating technique.  Sect. 3 is the core of this note: We define the minimal set of charges which suffices to produce the most general solution modulo the action of $G_{(3)}$; We then motivate the role of the STU model as a universal subsector of $\mathcal{N}>2$ or rank-3 symmetric $\mathcal{N}=2$ supergravities, in the sense described above; Eventually we  characterize the physical properties
of the non-extremal Harrison-transformed solution through the matrices $Q$ and $Q_\psi$ and define the limits which yield BPS (static and rotating singular) and non-BPS under-rotating solutions. This unifying geometric procedure includes previously found limits to BPS solutions \cite{Cvetic:1996kv}.  Finally, in Sect. 4 we derive the explicit form of the interpolating solution in the $T^3$ model and work out the extremal limits. The complete non-extremal $D=4$ solution is written for both the minimal sets of charges $(q_0,\,p^1)$ and $(p^0,\,q_1)$ of the model, including the fully-integrated vector potentials. Such complete $D=4$ description of these solutions, to our knowledge, was not present in the literature.\par
In the final stage of preparation of the manuscript, the work \cite{Cvetic:2013cja} came out, in which the $D=3$ description of the non-extremal solution in the STU model is given.

\section{The General Setting}\label{section1}
In this section we define the mathematical setting and review the solution-generating technique.
We consider stationary solutions in an extended, ungauged  $D=4$ supergravity with $n_s$ scalar and $n_v$ vector fields. We shall use the notations of \cite{Chemissany:2012nb,Andrianopoli:2012ee}.
The scalar fields $\phi^s$ parametrize a homogeneous, symmetric scalar manifold of the form:
\begin{equation}
\mathcal{M}^{(D=4)}_{scal}=\frac{G_4}{H_4}\,,
\end{equation}
$G_4$ being the semisimple isometry group and $H_4$ its maximal compact subgroup. The group $G_4$ also defines the global on-shell symmetry of the theory through its combined action on the scalar fields and on the vector field strengths and their magnetic duals as an electric-magnetic duality group.
The $D=4$ stationary metric we start from has the general form:
\begin{equation}
ds^2=-e^{2U}\,(dt+\omega)^2+e^{-2U}\,g^{(3)}_{ij}\,dx^i\,dx^j\,\,\,,\,\,\,i,j=1,2,3\,,\label{statmet}
\end{equation}
where $\omega=\omega_i\,dx^i$, $i=1,2,3$. \par
Upon formally reducing to three dimensions along the time direction and dualizing the vector fields to scalar fields, according to the prescription of \cite{Breitenlohner:1987dg}, we end up with a sigma model coupled to gravity, whose Lagrangian has the form:
\begin{align} \label{geodaction}
\frac{1}{\sqrt{g^{(3)}}}\,\mathcal{L}_{(3)} &= \frac{1}{2}\,\cR - \tfrac{1}{2}
G_{ab}(z)\partial_i{z}^a\partial^i{z}^b=\nn\\
&=\frac{1}{2}\, \cR -[ \partial_i U \partial^i
U+\tfrac{1}{2}\,G_{rs}\,\partial_i{\phi}^r\,\partial^i{\phi}^s +
\tfrac{1}{2}\e^{-2\,U}\,\partial_i{\mathcal{Z}}^T\,\mathcal{M}_{(4)}\,\partial^i{\mathcal{Z}} +\nn\\ &+
\tfrac{1}{4}\e^{-4\,U}\,(\partial_i{a}+\mathcal{Z}^T\mathbb{C}\partial_i{\mathcal{Z}})(\partial^i{a}+\mathcal{Z}^T
\mathbb{C}\partial^i{\mathcal{Z}})]\,,
\end{align}
where $g^{(3)}\equiv {\rm det}(g^{(3)})$.
Here, all the propagating degrees of freedom have been reduced to scalars by 3D Hodge-dualization.
In particular, the scalars $\mathcal{Z} =
(\mathcal{Z}^\Lambda,\mathcal{Z}_\Lambda)= \{\mathcal{Z}^M\}$ include the electric components $A^\Lambda_0$ of
the 4D vector fields together with the Hodge dual of their magnetic components $A^\Lambda_i$ ($i=1,2,3$) and $a$
is related to the Hodge-dual of the 3D Kaluza-Klein vector $\omega_i$. More precisely,
\begin{eqnarray}
A^\Lambda_{(4)}&=& A^\Lambda_0 (dt + \omega)+ A^\Lambda_{(3)} \,,\quad A^\Lambda_{(3)}\equiv A^\Lambda_i dx^i\,,\\
\mathbb{F}^M&=&\left(
                       \begin{array}{c}
                         F^\Lambda_{(4)} \\
                         \cG_{\Lambda (4)} \\
                       \end{array}
                     \right)
                     =d \mathcal{Z}^M \wedge (dt + \omega) + e^{-2 U}\mathbb{C}^{MN}\mathcal{M}_{(4) NP} {}^{*_3} d\mathcal{Z}^P\,,\\
da&=& - e^{4 U} \,{}^{*_3}d\omega - \mathcal{Z}^T \mathbb{C} d\mathcal{Z}\,,
\end{eqnarray}
where $F^\Lambda_{(4)}= dA^\Lambda_{(4)}$, $\cG_{\Lambda (4)} =-\frac{1}{2} {}^*\left(\frac{\partial \cL}{\partial
F^\Lambda_{(4)}}\right)$,  $\cM_{(4)}(\phi)$ is the negative-definite symmetric, symplectic matrix depending on 4D scalar fields and ${}^{*_3}$ denotes the Hodge-duality operation in $D=3$. The symplectic vector $\mathbb{F}^M$ transforms under the duality action of $G_4$ in a symplectic representation, ${\bf R}$.
 The scalar fields of the $D=3$  sigma-model, $(\phi^I)\equiv \{U,a,\phi^s,\mathcal{Z}^M\}$, span a homogeneous-symmetric, pseudo-Riemannian  scalar manifold $\mathcal{M}_{scal}$ of the form
\begin{equation}
\mathcal{M}_{scal}=\frac{G_{(3)}}{H^*}\,.
\end{equation}
The isometry group $G_{(3)}$ of the target space is the global symmetry group of $\mathcal{L}_{(3)}$
and $H^*$ is a suitable non-compact semisimple maximal subgroup of it. We refer the reader to Appendix \ref{A1} for the details about the geometry of $\mathcal{M}_{scal}$.\par
The scalar fields $\phi^I$ define a local \emph{solvable parametrization} of the coset (see Appendix \ref{A1}), and the coset representative  is chosen to be
\begin{equation}
\mathbb{L}(\phi^I)=\exp(-a T_\bullet)\,\exp(\sqrt{2}
\mathcal{Z}^M\,T_M)\,\exp(\phi^r\,T_r)\,\exp(2\,U
H_0)\,,\label{cosetr3}
\end{equation}
where $T_{\mathcal{A}}=\{H_0,\,T_\bullet,\,T_s,\,T_M\}$ are the solvable generators defined in  Appendix \ref{A1}.
Since the generators $T_M$ transform under the adjoint action of $G_4\subset G_{(3)}$ in the symplectic duality representation ${\bf R}$ of the electric-magnetic charges, we shall use for them the following notation: $(T_M)=(T_{q_\Lambda},\,T_{p^\Lambda})$.\par
All the formulas related to the group $G_{(3)}$ and its Lie algebra $\mathfrak{g}$ are referred to a matrix representation in which the subalgebra $\mathfrak{H}^*$ of $\mathfrak{g}$, Lie algebra of $H^*$, and its orthogonal complement $\mathfrak{K}^*$ are defined by a pseudo-Cartan involution $\sigma$ acting on a matrix $M$ as $\sigma(M)=- \eta M^\dagger\eta$, where $\eta$ is a suitable $H^*$-invariant metric (see appendix \ref{A1}).\par
It is also useful to introduce the
hermitian, $H^*$-invariant matrix $\mathcal{M}$ which, in a chosen matrix representation, reads:
\begin{equation}
\mathcal{M}(\phi^I)\equiv \mathbb{L}\eta
\mathbb{L}^\dagger=\mathcal{M}^\dagger\,.\label{cm}
\end{equation}
In terms of $\mathcal{M}$ we can write the $D=3$ Noether-current associated with a stationary solution $\phi^I(x^i)$:
\begin{equation}
J_i\equiv \frac{1}{2}\partial_i \phi^I\,\mathcal{M}^{-1}\partial_I
\mathcal{M}\,.\label{curr}
\end{equation}
The $\mathfrak{g}$-valued Noether-charge matrix reads:
\begin{equation}
Q=\frac{1}{4\pi}\int_{S_2} {}^{*_{3}}J=\frac{1}{4\pi}\,\int \sqrt{g^{(3)}}\,
J^r d\theta d\varphi \,,
\end{equation}
the index of $J$ being raised using $g^{(3)\,ij}$. \par
Restricting to axisymmetric solutions and denoting by $\psi=\partial_\varphi$ the angular Killing vector,
all the fields will  only depend on $x^m=(r,\theta)$. It is useful to describe the global rotation of the solution by means of a new $\mathfrak{g}$-valued matrix $Q_\psi$, first defined in \cite{Andrianopoli:2012ee} as:
\begin{equation}
Q_\psi =-\frac{3}{4\pi}\,\int_{S_2^\infty}\psi_{[i}\,J_{
j]}\,dx^i\wedge dx^j=\frac{3}{8\pi}\,\int_{S_2^\infty} g^{(3)}_{\varphi\varphi}\,J_\theta\,d\theta d\varphi\,.
\end{equation}
The ADM-mass, NUT-charge, electric and magnetic charges $\Gamma^M=(p^\Lambda,\,q_\Lambda)$, scalar charges $\Sigma_s$ and angular momentum $M_\varphi$, associated with the solution are then obtained as components of $Q$ and $Q_\psi$:
\begin{align}
M_{ADM}&=k\,{\rm
Tr}(H_0^\dagger\,Q)\,,\,\,\,n_{NUT}=-k\,{\rm
Tr}(T_\bullet^\dagger\,Q)\,,\,\,\,\Gamma^M=\sqrt{2}\,k\,\mathbb{C}^{MN}\,{\rm
Tr}(T_N^\dagger\,Q)\,,\,\,\,\Sigma_s=k\,{\rm
Tr}(T_s^\dagger\,Q)\nonumber\\
M_\varphi&=k\,{\rm Tr}(T_\bullet^\dagger\,Q_\psi)\,,
\end{align}
Being $G_{(3)}$ the global symmetry group of the $D=3$ theory, its action on any solution $\phi^I(x^i)$ yields another solution $\phi^{\prime I}(x^i)$, related to the original one through the matrix equation:
\begin{equation}
\mathcal{M}(\phi^{\prime I}(x^i))=g\,\mathcal{M}(\phi^{I}(x^i))\,g^\dagger\,.
\end{equation}
If a solution is defined by a unique point at radial infinity $\phi_0=(\phi_0^I)$, $\phi^I_0=\lim_{r\rightarrow \infty} \phi^I$, we can always map $\phi_0$ into the origin $O$ of the manifold, defined by $\phi^I_0\equiv 0$, by means of a transformation in $G_{(3)}/H^*$. Thus when studying the properties of these solutions with respect to the action of $G_{(3)}$, without loss of generality we can fix the point at infinity to coincide with the origin $\phi_0\equiv O$. This choice breaks $G_{(3)}$ to the isotropy group $H^*$. As a consequence of
this the two matrices $Q,\,Q_\psi$ always lie in the coset space $\mathfrak{K}^*$.
\par
Among the $H^*$-transformations, a special role is played by the Harrison transformations, which are generated by the non-compact generators $\mathbb{J}_M$ of $\mathfrak{H}^*$:
\begin{equation}
\mathbb{J}_M\equiv \frac{1}{2}\,(T_M+T_M^\dagger)\,.\label{JM}
\end{equation}
The space $\mathbb{J}^{(R)}$ spanned by $\mathbb{J}_M$ is the carrier of a representation ${\bf R}$ with respect to the adjoint action of the maximal compact subgroup $H_c$ of $H^*$.\footnote{With an abuse of notation, we use the same symbol ${\bf R}$ to denote the symplectic duality representation of $G_4$ and the corresponding representation of $H_c$, both being related to the electric and magnetic charges.} This group has the general form $H_c={\rm U}(1)_E\times H_4$, ${\rm U}(1)_E$ being contained in the Ehlers group ${\rm SL}(2,\mathbb{R})_E$.\par
We also define the subspace $\mathbb{K}^{(R)}$ of the coset space $\mathfrak{K}^*$ spanning the negative-signature directions of the metric, and defining, just as   $\mathbb{J}^{(R)}$, the support of a representation ${\bf R}$ of $H_c$. Its compact generators have, in the chosen matrix representation, the form:
\begin{equation}
\mathbb{K}_M\equiv \frac{1}{2}\,(T_M-T_M^\dagger)\,.\label{KM}
\end{equation}

\section{The Symmetric Models}\label{section2}
We consider in this Section the solution obtained by acting on the non-extremal Kerr black hole by means of the
maximal number of commuting Harrison transformations. These are generated by the maximal abelian subspace $\mathbb{J}^{(N)}$ of
$\mathbb{J}^{(R)}$.  From general group theoretical arguments, it follows that the dimension of such space is nothing but the rank of the coset $H^*/H_c$:
\begin{equation}
p={\rm dim}(\mathbb{J}^{(N)})={\rm rank}\left(\frac{H^*}{H_c}\right)\,.\label{p}
\end{equation}
This number coincides with the number of parameters of the normal form of a generic vector in the symplectic  representation ${\bf R}$ (i.e. the representation of the electric and magnetic charges), with respect to the action of $H_c$. By this we mean that $\mathbb{J}^{(N)}$ is the minimal subspace in which  a generic representative, $\xi^M \mathbb{J}_M$, of $\mathbb{J}^{(R)}$ can be rotated by means of an $H_c$  transformation.
Let us recall that $H_c = U(1)_E\times H_4$, so that the total number of parameters of the normal form of the same vector with respect to the action of $H_4$ alone is $p+1$, i.e. the $D=4$ seed solution, with respect to the action of $H_4$ is a $(p+1)$-parameter solution.\par
The above discussion also applies to the space of compact generators $\mathbb{K}^{(R)}$ in the coset, for which we can define a normal subspace $\mathbb{K}^{(N)}$ of dimension $p$. This proves that the charges of the most general solution can always be reduced in number to $p$ by means of the action of the global symmetry group in $D=3$ \cite{Bergshoeff:2008be}. \par
In the maximal supergravity, for example, $p={\rm rank}\left(\frac{{\rm SO}^*(16)}{{\rm U}(8)}\right)=4$, the same being true for the half-maximal theory, $p={\rm rank}\left(\frac{{\rm SO}(6,2)\times {\rm SO}(2,6+n)}{{\rm SO}(2)^2\times {\rm SO}(6)\times {\rm SO}(6+n)}\right)=4$, and for the $\mathcal{N}=2$ symmetric models with rank-3 scalar manifold in $D=4$ (for this class of theories, $p=$ rank $+1$). The simplest representative of the latter class of models is the $STU$ one, which is a consistent truncation of all the others, besides being a truncation of the maximal and half-maximal theories. Therefore the space $\mathbb{J}^{(N)}$ is contained in the spaces of Harrison generators $\mathbb{J}^{(R)}$ of all the above mentioned symmetric models, and thus, for the sake of simplicity, we can restrict ourselves to the simplest $STU$ model and act on its Kerr solution by means of a transformation generated by $\mathbb{J}^{(N)}$.
As far as the $\mathcal{N}=2$ theories are concerned, we also have lower rank models such as the  rank-2 model $ST^2$, with $p=3$, and the rank-1 $T^3$-model with $p=1$, which we shall be dealing with in next section.\footnote{There are also the $\mathcal{N}=2$ models with rank-1 special K\"ahler scalar manifold of the form ${\rm U}(1,n)/[{\rm U}\times {\rm U}(n)]$, which we are not going to deal with here. }\par
We shall use the notations of \cite{Bergshoeff:2008be,Chemissany:2012nb,Andrianopoli:2012ee}. Let us denote by $\mathcal{J}_\ell$ the generators of $\mathbb{J}^{(N)}$ and $\mathcal{K}_\ell$ those of $\mathbb{K}^{(N)}$. These two spaces define, together with the generators $\mathcal{H}_\ell=-[\mathcal{J}_\ell,\,\mathcal{K}_\ell]$, a universal subgroup ${\rm SL}(2,\mathbb{R})^p$, each factor being generated by the triple $\{\mathcal{K}_\ell,\,\mathcal{J}_\ell,\,\mathcal{H}_\ell\}$, $\ell=0,\dots, p-1$.\par
% If we denote by $(T_M)=(T_{q_\Lambda},\,T_{p^\Lambda})$ the nilpotent generators parametrized by the scalars $(\mathcal{Z}^M)=(\mathcal{Z}^\Lambda,\,\mathcal{Z}_\Lambda)$ of the $D=3$ sigma model, originating from the $D=4$ vector fields, then, in the chosen matrix representation, we can write the generators of $\mathbb{K}^{(R)}={\rm Span}(\mathbb{K}_M)$ and $\mathbb{J}^{(R)}={\rm Span}(\mathbb{J}_M)$ as follows:
%\begin{equation}
%\mathbb{K}_M=\frac{T_M-T_M^\dagger}{2}\,\,;\,\,\,\,\mathbb{J}_M=\frac{T_M+T_M^\dagger}{2}\,.
%\end{equation}
If the $D=4$ theory results from the dimensional reduction of a $D=5$ one on  a circle, then the field strengths and their magnetic duals transform, with respect to the action of the $D=4$ duality group $G_4$, in a specific symplectic frame. In this frame, which in the $\mathcal{N}=2$ theories is the special-coordinate one, there are  two spaces $\mathbb{J}^{(N)}$ (and thus $\mathbb{K}^{(N)}$) defined respectively by the generators $\{\mathcal{T}_\ell\}=\{T_{q_0},\,T_{p^i}\}$ or  $\{\mathcal{T}_\ell\}=\{T_{p^0},\,T_{q_i}\}$ \cite{Bergshoeff:2008be}:
\begin{equation}
\mathcal{K}_\ell=\frac{\mathcal{T}_\ell-\mathcal{T}_\ell^\dagger}{2}\,\,;\,\,\,\,\mathcal{J}_\ell=
\frac{\mathcal{T}_\ell+\mathcal{T}_\ell^\dagger}{2}\,.
\end{equation}
We shall distinguish the second set of charges from the former one by means of a prime on the corresponding generators.
The two sets of charges $\{q_0,\,p^i\}$ and $\{p^0,\,q_i\}$ define indeed the two normal forms of the electric-magnetic charge vector with respect to the action of $H_c$.\par
Consider now the  Harrison transformations:
\begin{equation}
\mathcal{O}_{(q_0,\,p^i)}=e^{\log(\beta_\ell) \mathcal{J}_\ell}\,\,\,;\,\,\,\,\mathcal{O}_{(p^0,\,q_i)}=e^{\log(\beta_\ell) \mathcal{J}'_\ell}\,,\label{Harris}
\end{equation}
and act by means of them on the non-extremal Kerr solution.
If $\mathbb{L}(\phi^I)$ is the coset representative of the $D=3$ sigma model, and $\mathcal{M}(\phi^I)=\mathbb{L}(\phi^I)\eta \mathbb{L}(\phi^I)^\dagger$, the new $D=3$ solution $\phi^I(r,\theta)$ is obtained by solving the matrix equation:
\begin{equation}
\mathcal{M}(\phi^I(r,\theta))=\mathcal{O}\mathcal{M}(\phi_K^I(r,\theta))\mathcal{O}^{\dagger}\,,\label{mateq}
\end{equation}
where $\mathcal{O}$ generically denotes any of the two transformations in (\ref{Harris}), while $\phi_K^I(r,\theta)$ are the $D=3$ scalar fields defining the Kerr solution. In the next section we shall work out explicitly this solution in the $T^3$ model and derive the full $D=4$ one for the two normal sets of charges. Here we shall make a general analysis by  considering the effect of the above transformations on the charges and defining the relevant extremal limits. Although we consider here the $STU$ model,  the analysis applies to  lower rank models as well, as we shall illustrate in the next section. In the following we shall also use a real matrix representation of $G_{(3)}$, in which the Harrison generators are symmetric, so that $\mathcal{O}^\dagger=\mathcal{O}^T=\mathcal{O}$.\par
The Noether-charge matrix $Q$ and the matrix $Q_\psi$ associated with the rotation of the solution, for the Kerr black hole read \cite{Andrianopoli:2012ee}:
\begin{equation}
Q^{(K)}=2\,m\,H_0\,\,,\,\,\,Q^{(K)}_\psi=2\,m\,\alpha\,T_\bullet\,.
\end{equation}
The corresponding matrices for the transformed solution are readily evaluated from (\ref{mateq}):
\begin{equation}
Q=\mathcal{O}^{-1}\,Q^{(K)}\,\mathcal{O}\,\,;\,\,\,\,Q_\psi=\mathcal{O}^{-1}\,Q^{(K)}_\psi\,\mathcal{O}\,.\label{harq}
\end{equation}
To compute the new matrices it is useful to express $Q^{(K)}$ and $Q^{(K)}_\psi$ in terms of eigenmatrices of the adjoint action of  $\mathcal{J}_\ell$ on $\mathfrak{K}^*$. These have the form
$\mathcal{N}^{\vec{w}}$, $\vec{w}$ being a subset of the roots of $G_{(3)}$ such that:
\begin{equation}
[\mathcal{J}_\ell,\,\mathcal{N}^{\vec{w}}]=w_\ell\,\mathcal{N}^{\vec{w}}\,.
\end{equation}
The advantage of this notation is that a boost generated by the $\mathcal{J}_\ell$'s amounts to a rescaling of these nilpotent matrices:
\begin{equation}
e^{-\log(\beta_\ell)\,\mathcal{J}_\ell}\mathcal{N}^{\vec{w}}e^{\log(\beta_\ell)\,\mathcal{J}_\ell}=
\frac{1}{\beta_\ell^{w_\ell}}\,\mathcal{N}^{\vec{w}}\,.\label{resc}
\end{equation}
As far as the $STU$ model is concerned ($p=4$) we have:
\begin{equation}
\vec{w}=\left\{\{\pm 1,0,0,0\},\,\{0,\pm 1,0,0\},\,\{0,0,\pm 1,0\},\,\{0,0,0,\pm 1\},\,\frac{1}{2}\{\epsilon_0,\,\epsilon_1,\,\epsilon_2,\,\epsilon_3\}\right\}\, ,
\end{equation}
where $\epsilon_\ell=\pm 1$  ($\ell=0,1,2,3$).
As far as the last roots $\vec{w}=\frac{1}{2}\{\epsilon_0,\,\epsilon_1,\,\epsilon_2,\,\epsilon_3\}$ are concerned, they are eight since, depending on the chosen normal form, the following constraints on the signatures hold:
\begin{equation}
\begin{cases}\prod_{\ell=0}^3\epsilon_\ell=1,\,\, &(q_0,\,p^i )\cr \prod_{\ell=0}^3\epsilon_\ell=-1,\, \, &(p^0,\,q_i)\,.\end{cases}
\end{equation}
For the sake of notational simplicity we shall define $\mathcal{N}^{\epsilon_\ell}_\ell\equiv \mathcal{N}^{(0,\dots,\epsilon_\ell,\dots,0)}$.
One can verify the general properties:\footnote{Below we use the notation $\sum_{\prod\epsilon=\pm 1}$ to indicate the sum over all values of $\epsilon_\ell=\pm 1$, constrained by the condition $\prod_{\ell=0}^3\epsilon_\ell=\pm 1$.}
\begin{align}
Q^{(K)}&=\frac{m}{2}\,\sum_{\ell=0}^3\sum_{\epsilon_\ell=\pm 1} \mathcal{N}^{\epsilon_\ell}_\ell\,,\nonumber\\
Q^{(K)}_\psi &=\begin{cases}\frac{m\,\alpha}{2}\,\sum_{\prod\epsilon=1}\,
\mathcal{N}^{\frac{1}{2}(\epsilon_0,\epsilon_1,\epsilon_2,\epsilon_3)},\,\, (q_0,\,p^i )\,\mbox{-case}\cr \frac{m\,\alpha}{2}\,\sum_{\prod\epsilon=-1}\,
\mathcal{N}^{\frac{1}{2}(\epsilon_0,\epsilon_1,\epsilon_2,\epsilon_3)},\,\, (p^0,\,q_i )\,\mbox{-case}\end{cases} \label{qkn}
\end{align}
Using Eq.s (\ref{harq}), (\ref{resc}), (\ref{qkn}) we can compute the transformed charge-matrices:
\begin{align}
Q&=\frac{m}{2}\,\sum_{\ell=0}^3\sum_{\epsilon_\ell=\pm 1} \frac{1}{\beta_\ell^{\epsilon_\ell }}\mathcal{N}^{\epsilon_\ell}_\ell\,,\nonumber\\
Q_\psi &=\frac{m\,\alpha}{2}\,\sum_{\prod\epsilon=1}\,\frac{1}{\prod_\ell \beta_\ell^{\frac{\epsilon_\ell}{2}}}
\mathcal{N}^{\frac{1}{2}(\epsilon_0,\epsilon_1,\epsilon_2,\epsilon_3)},\,\qquad (q_0,\,p^i )\,\mbox{-case}\,,\nonumber\\
Q_\psi &=\frac{m\,\alpha}{2}\,\sum_{\prod\epsilon=-1}\,\frac{1}{\prod_\ell \beta_\ell^{\frac{\epsilon_\ell}{2}}}
\mathcal{N}^{\frac{1}{2}(\epsilon_0,\epsilon_1,\epsilon_2,\epsilon_3)},\,\qquad (p^0,\,q_i )\,\mbox{-case}\,.\nonumber
\end{align}
The ADM mass, the electric and magnetic charges and angular momentum are computed from $Q$ and $Q_\psi$ to be:
\paragraph{$(q_0,\,p^i)$-case.}
\begin{align}
M_{ADM}&=\frac{m}{8}\,\sum_{\ell=0}^3\sum_{\epsilon_\ell=\pm 1}\frac{1}{\beta_\ell^{\epsilon_\ell }}\,\,,\,\,\,
q_0=\frac{m}{2\sqrt{2}}\sum_{\epsilon_0=\pm 1} \frac{\epsilon_0}{\beta_0^{\epsilon_0 }}\,,\,\,\,
p^i=-\frac{m}{2\sqrt{2}}\sum_{\epsilon_i=\pm 1} \frac{\epsilon_i}{\beta_i^{\epsilon_i }}\,\nonumber\\
M_\varphi&=\frac{m\,\alpha}{8}\,\sum_{\prod\epsilon=1}\,\frac{1}{\prod_\ell \beta_\ell^{\frac{\epsilon_\ell}{2}}}\,.
\end{align}
The evaluation of the quartic $G_4$-invariant of the representation ${\bf R}$ of the electric and magnetic charges gives:
\begin{equation}
I_4(p,q)=4\,q_0\,p^1p^2p^3\,.
\end{equation}
\paragraph{$(p^0,\,q_i)$-case.}
\begin{align}
M_{ADM}&=\frac{m}{8}\,\sum_{\ell=0}^3\sum_{\epsilon_\ell=\pm 1}\frac{1}{\beta_\ell^{\epsilon_\ell }}\,\,,\,\,\,
p^0=-\frac{m}{2\sqrt{2}}\sum_{\epsilon_0=\pm 1} \frac{\epsilon_0}{\beta_0^{\epsilon_0 }}\,,\,\,\,
q_i=\frac{m}{2\sqrt{2}}\sum_{\epsilon_i=\pm 1} \frac{\epsilon_i}{\beta_i^{\epsilon_i }}\,\nonumber\\
M_\varphi&=\frac{m\,\alpha}{8}\,\sum_{\prod\epsilon=-1}\,\frac{1}{\prod_\ell \beta_\ell^{\frac{\epsilon_\ell}{2}}}\,.
\end{align}
the quartic invariant reads in this case:
\begin{equation}
I_4(p,q)=-4\,p^0\,q_1q_2q_3\,.
\end{equation}
\subsection{Extremal Limits}
Next we discuss singular limits of the above solutions effected by sending the Harrison parameters $\beta_\ell$ to zero or infinity and, at the same time, the Kerr parameters $m$ and $\alpha$ to zero. This limiting procedure extends to symmetric supergravities the procedure found by Rasheed and Larsen  for the Kaluza-Klein theory.\par
Let us start redefining:
\begin{equation}
\beta_\ell=m^{\sigma_\ell}\,\alpha_\ell\,\,;\,\,\,\alpha=m\,\Omega\,,
\end{equation}
where $\sigma_\ell=\pm 1$. Next we send $m$ to zero by keeping $\alpha_\ell$ and $\Omega$ fixed. As we illustrate below, the effect of this limit is to make the matrices $Q$ and $Q_\psi$ nilpotent, thus making the resulting solution extremal. Depending on the choice of $\sigma_\ell$, we can recover in the limit all the $G_4$ orbits of the electric and magnetic charges. In particular, as far as the regular extremal solutions are concerned, these orbits are\cite{Bellucci:2006xz}:
\begin{align}
\mbox{BPS}&:\,\,I_4(p,q)>0\,\,\,\,\mbox{$\mathbb{Z}_3$-symmetry on the  $p^i$ and the $q_i$}\,,\nonumber\\
\mbox{non-BPS}_1&:\,\,I_4(p,q)>0\,\,\,\,\mbox{no $\mathbb{Z}_3$-symmetry}\,,\nonumber\\
\mbox{non-BPS}_2&:\,\,I_4(p,q)<0\,.\nonumber
\end{align}
Let us consider the two normal forms separately.
\begin{itemize}
\item  {\bf $(q_0,\,p^i)$-case.}

 The rescaling produces the following expressions of the relevant physical quantities:
\begin{align}
M_{ADM}&=\frac{1}{8}\,\sum_{\ell=0}^3\sum_{\epsilon_\ell=\pm 1}\frac{m^{1-\sigma_\ell\epsilon_\ell}}{\alpha_\ell^{\epsilon_\ell }}\,\,,\,\,\,
q_0=\frac{1}{2\sqrt{2}}\sum_{\epsilon_0=\pm 1} \frac{\epsilon_0\,m^{1-\sigma_0\epsilon_0}}{\alpha_0^{\epsilon_0 }}\,,\,\,\,
p^i=-\frac{1}{2\sqrt{2}}\sum_{\epsilon_i=\pm 1} \frac{\epsilon_i\,m^{1-\sigma_i\epsilon_i}}{\alpha_i^{\epsilon_i }}\,,\nonumber\\
M_\varphi&=\frac{\Omega}{8}\,\sum_{\prod\epsilon=1}\,\frac{m^{2-\frac{1}{2}\sum_\ell\sigma_\ell\epsilon_\ell}}{\prod_\ell \alpha_\ell^{\frac{\epsilon_\ell}{2}}}\,.
\end{align}
Notice that, in the limit $m\rightarrow 0$, the only terms surviving are those for which $\epsilon_\ell=\sigma_\ell$ and the resulting values of the  ADM mass and the electric and magnetic charges read:
\begin{equation}
M_{ADM}=\frac{1}{8}\,\sum_{\ell=0}^3\frac{1}{\alpha_\ell^{\sigma_\ell }}\,\,,\,\,\,
q_0=\frac{1}{2\sqrt{2}} \frac{\sigma_0}{\alpha_0^{\sigma_0 }}\,,\,\,\,
p^i=-\frac{1}{2\sqrt{2}}\frac{\sigma_i}{\alpha_i^{\sigma_i }}\,
\end{equation}
The value of the quartic invariant is:
\begin{equation}
I_4(p,q)=\frac{1}{16\,\prod_\ell \alpha_\ell^{\sigma_\ell}}\,\left(-\prod_\ell \sigma_\ell\right)\,.
\end{equation}
On the other hand $M_\varphi$ contains non-vanishing terms in the $m\rightarrow 0$ limit, only if $\prod_\ell \sigma_\ell=1$ (i.e. $I_4<0$). As a consequence of this, only the $\mbox{non-BPS}_2$ solution has a residual angular momentum. This is the \emph{under-rotating} (single center) solution discussed in  \cite{Bena:2009ev}. In this case the
angular momentum reads:
\begin{equation}
M_\varphi=\frac{\Omega}{8\,\prod_\ell \alpha_\ell^{\frac{\sigma_\ell}{2}}}=\frac{\Omega}{2}\,\sqrt{|I_4(p,q)|}\,.
\end{equation}
Summarizing we find:
\begin{align}
\mbox{BPS}&:\,\,\prod_\ell \sigma_\ell=-1\,,\,\,\,\sigma_1=\sigma_2=\sigma_3\,\,,\,\,\,M_\varphi=0\,,\nonumber\\
\mbox{non-BPS}_1&:\,\,\prod_\ell \sigma_\ell=-1\,,\,\,\,\sigma_i\,\,\mbox{not all equal}\,\,,\,\,\,M_\varphi=0\,,\nonumber\\
\mbox{non-BPS}_2&:\,\,\prod_\ell \sigma_\ell=1\,,\,\,\,M_\varphi\neq 0\,.\nonumber
\end{align}
\item {\bf $(p^0,\,q_i)$-case.}

The rescaling produces the following expressions of the relevant physical quantities:
\begin{align}
M_{ADM}&=\frac{1}{8}\,\sum_{\ell=0}^3\sum_{\epsilon_\ell=\pm 1}\frac{m^{1-\sigma_\ell\epsilon_\ell}}{\alpha_\ell^{\epsilon_\ell }}\,\,,\,\,\,
p^0=-\frac{1}{2\sqrt{2}}\sum_{\epsilon_0=\pm 1} \frac{\epsilon_0\,m^{1-\sigma_0\epsilon_0}}{\alpha_0^{\epsilon_0 }}\,,\,\,\,
q_i=\frac{1}{2\sqrt{2}}\sum_{\epsilon_i=\pm 1} \frac{\epsilon_i\,m^{1-\sigma_i\epsilon_i}}{\alpha_i^{\epsilon_i }}\,,\nonumber\\
M_\varphi&=\frac{\Omega}{8}\,\sum_{\prod\epsilon=-1}\,\frac{m^{2-\frac{1}{2}\sum_\ell\sigma_\ell\epsilon_\ell}}{\prod_\ell \alpha_\ell^{\frac{\epsilon_\ell}{2}}}\,.
\end{align}
As in the previous case, in the limit $m\rightarrow 0$, the only terms surviving are those for which $\epsilon_\ell=\sigma_\ell$ and the resulting values of the  ADM mass and the electric and magnetic charges read:
\begin{equation}
M_{ADM}=\frac{1}{8}\,\sum_{\ell=0}^3\frac{1}{\alpha_\ell^{\sigma_\ell }}\,\,,\,\,\,
q_0=-\frac{1}{2\sqrt{2}} \frac{\sigma_0}{\alpha_0^{\sigma_0 }}\,,\,\,\,
p^i=\frac{1}{2\sqrt{2}}\frac{\sigma_i}{\alpha_i^{\sigma_i }}\,.
\end{equation}
The value of the quartic invariant is:
\begin{equation}
I_4(p,q)=\frac{1}{16\,\prod_\ell \alpha_\ell^{\sigma_\ell}}\,\left(\prod_\ell \sigma_\ell\right)\,.
\end{equation}
Similarly to the $(q_0,p^i)$-case,  in the $m\rightarrow 0$ limit $M_\varphi$ contains non-vanishing terms only if $\prod_\ell \sigma_\ell=-1$ (i.e. $I_4<0$). Consequently only the $\mbox{non-BPS}_2$ solution has a residual angular momentum:
 \begin{equation}
M_\varphi=\frac{\Omega}{8\,\prod_\ell \alpha_\ell^{\frac{\sigma_\ell}{2}}}=\frac{\Omega}{2}\,\sqrt{|I_4(p,q)|}\,,
\end{equation}
 yielding the  \emph{under-rotating} (single center) solution discussed in  \cite{Bena:2009ev}.
Summarizing we find:
\begin{align}
\mbox{BPS}&:\,\,\prod_\ell \sigma_\ell=1\,,\,\,\,\sigma_1=\sigma_2=\sigma_3\,\,,\,\,\,M_\varphi=0\,,\nonumber\\
\mbox{non-BPS}_1&:\,\,\prod_\ell \sigma_\ell=1\,,\,\,\,\sigma_i\,\,\mbox{not all equal}\,\,,\,\,\,M_\varphi=0\,,\nonumber\\
\mbox{non-BPS}_2&:\,\,\prod_\ell \sigma_\ell=-1\,,\,\,\,M_\varphi\neq 0\,.\nonumber
\end{align}
\end{itemize}
In the two cases discussed above, in the $m\rightarrow 0$ limit both  the matrices $Q$ and $Q_\psi$ become nilpotent. In the $\mbox{non-BPS}_2$ case:
\begin{equation}
Q=\frac{1}{2}\,\sum_\ell \frac{1}{\alpha_\ell^{\sigma_\ell}}\,\mathcal{N}^{\sigma_\ell}_\ell\,,\,\,\,Q_\psi=
\frac{\Omega}{2\prod_\ell \alpha_\ell^{\frac{\sigma_\ell}{2}}}
\mathcal{N}^{\frac{1}{2}(\sigma_0,\sigma_1,\sigma_2,\sigma_3)}\,,
\end{equation}
$Q$, in the fundamental of $G_{(3)}$, is step-3, while $Q_\psi$ is step-2. In this case, the $G_{(3)}$-invariant regularity condition of the original Kerr solution:
\begin{equation}
m^2\ge \alpha^2\,\,\Leftrightarrow \,\,\,\,{\rm Tr}[Q^2]\ge \frac{2}{k}\,\frac{{\rm Tr}[Q_\psi^2]}{{\rm Tr}[Q^2]}\,,\label{regcond}
\end{equation}
is saturated in the $m\rightarrow 0$ limit since both sides vanish separately (the degree of nilpotency of $Q_\psi$, in the limit, being smaller than that of $Q$), and the resulting solution is extremal. For the BPS and the $\mbox{non-BPS}_1$ solutions $Q$ has the same expression as in eq. (3.25), while $Q_\psi=0$, the regularity condition being thus still saturated.\par
\vskip 5mm
We could perform a different singular limit, which coincides with the one described above except that we keep $\alpha=\Omega$ fixed. As far as the ADM mass and the electric and magnetic charges are concerned, the discussion is the same as for the previous limits. The angular momentum however reads now:
\begin{align}
M_\varphi&=\frac{\Omega}{8}\,\sum_{\prod\epsilon=\pm 1}\,\frac{m^{1-\frac{1}{2}\sum_\ell\sigma_\ell\epsilon_\ell}}{\prod_\ell \alpha_\ell^{\frac{\epsilon_\ell}{2}}}\,.
\end{align}
The only contributing terms are those for which $\sum_\ell\sigma_\ell\epsilon_\ell=2$. This excludes the $\mbox{non-BPS}_2$ charge orbit for which $M_\varphi$ diverges and leaves just the $\mbox{BPS}$, \cite{Cvetic:1996kv}, and $\mbox{non-BPS}_1$ orbits. In this limit the degree of nilpotency of the resulting matrices $Q$ and $Q_\psi$ is the same and the regularity condition (\ref{regcond}) is no longer satisfied, since the left-hand side vanishes while the right hand side remains finite. These are (singular) rotating $\mbox{BPS}$ and $\mbox{non-BPS}_1$ orbits, the former was first studied in \cite{Behrndt:1997ny}.\par
It is important to stress that the BPS, $\mbox{non-BPS}_1$ and $\mbox{non-BPS}_2$  single center solutions that we defined through the above limiting procedure, are \emph{seed solutions} with respect to the action of $G_{(3)}$ (in our case of $H^*$ since we fixed the point at infinity to coincide with the origin $O$), of the most general $D=4$ solution of each class.
Indeed the corresponding nilpotent $H^*$-orbits in $\mathfrak{K}^*$ of the Noether-charge matrix $Q$ have representatives in the coset space of the submanifold
\begin{eqnarray}\label{genman}
\left(\frac{{\rm SL}(2,\mathbb{R})}{{\rm SO}(1,1)}\right)^p\subset \frac{G_{(3)}}{H^*}\,,
\end{eqnarray}
generated by $\mathcal{H}_\ell,\,\mathcal{K}_\ell$, or, equivalently, by $\mathcal{N}^{\epsilon_\ell}_\ell=\mathcal{H}_\ell-\epsilon_\ell\,\mathcal{K}_\ell$ \cite{Bergshoeff:2008be,marioetal,Chemissany:2012nb,Fre:2011uy}.\par
In the next section we shall construct the explicit solution in the $T^3$ model and discuss its extremal limits.

\section{Example: The $T^3$-Model}
The $T^3$ model is an $\mathcal{N}=2$ supergravity coupled to one vector multiplet (see  Appendix \ref{t3}), originating from pure $D=5$ supergravity. The two real scalar fields $\phi^s=(y,\varphi)$ are combined in a single complex scalar $T=y-i\,e^{\varphi}$ spanning  the special K\"ahler manifold ${\rm SL}(2,\mathbb{R})/{\rm SO}(2)$.\par
Upon timelike reduction to $D=3$, the scalar fields $\Phi^I=\{U,\phi^s,\,\mathcal{Z}^M,\,a\}$ are local parameters of the  pseudo-Riemannian coset space:
  \begin{equation}
  \mathcal{M}^{(D=3)}_{scal}=\frac{G_{(3)}}{H^*}=\frac{{\rm G}_{2(2)}}{{\rm SL}(2,\mathbb{R})^2}\,.
  \end{equation}
The mathematical details about the algebra $\mathfrak{g}_{2(2)}$ and the solvable generators $T_{\mathcal{A}}$, parametrized by the scalar fields $\phi^I$, are given in Appendix \ref{A2}, together with the precise definition of the coset representative.

\subsection{The Non-Extremal Solution}
In this model the normal form of the representation ${\bf R}$ of the electric and magnetic charges with respect to $H_c={\rm U}(1)_E\times {\rm U}(1)$ has $p=2$ parameters, according to the characterization in Eq. (\ref{p}). It can be characterized either by the set $(q_0,\,p^1)$ or by $(p^0,\,q_1)$. The corresponding Harrison generators will be denoted by $\mathcal{J}_\ell$ and $\mathcal{J}'_\ell$  respectively ($\ell=0,1$). The mathematical details are given in Appendix \ref{A2}.\par
In this subsection we first study the physical quantities associated with the Harrison-transformed solution. We shall restrict here, for the sake of simplicity, to the $(q_0,\,p^1)$ set of charges, keeping in mind that the corresponding analysis for the $(p^0,\,q_1)$-case is analogous. Later, in Section \ref{solution}, we shall present the full $D=4$ solutions corresponding to both sets of charges.\par
Along the lines of our discussion in the previous section, to study the effect of the Harrison transformations on the matrices $Q,\,Q_\psi$, we express these generators in terms of eigenmatrices with respect to the adjoint action of  $\mathcal{J}_\ell$, which can be chosen as a basis of Cartan generators of $\mathfrak{g}_{2(2)}$. We define a set of nilpotent generators $\mathcal{N}_{(a,b)}$ such that:
\begin{equation}
[\mathcal{J}_0,\,\mathcal{N}_{(a,b)}]=a\,\mathcal{N}_{(a,b)}\,,\,\,\,[\mathcal{J}_1,\,\mathcal{N}_{(a,b)}]=b\,\mathcal{N}_{(a,b)}\,,
\end{equation}
the coset space $\mathfrak{K}^*$ is spanned by the following
generators:
\begin{equation}
\mathfrak{K}^*=\mbox{Span}(\mathcal{N}_{\left(\frac{1}{2},\,\frac{3}{2}\right)},\,\mathcal{N}_{\left(\frac{1}{2},\,-\frac{1}{2}\right)},\,
\mathcal{N}_{(1,\,0)},\,\mathcal{N}_{(0,\,1)},\,\mathcal{N}_{\left(-\frac{1}{2},\,-\frac{3}{2}\right)},\,\mathcal{N}_{\left(-
\frac{1}{2},\,\frac{1}{2}\right)},\,
\mathcal{N}_{(-1,\,0)},\,\mathcal{N}_{(0,\,-1)})\,,\label{Nbasis}
\end{equation}
We refer the reader to Appendix \ref{A22} for the explicit form of these generators.\par
 Let us now consider the Kerr
solution  with $Q^{(K)}=2\,m\,H_0$ and $Q^{(K)}_\psi=2\,\alpha
\,m\,\mathbb{K}_\bullet$ and apply to it the transformation
$\mathcal{O}_{(q_0,\,p^1)}$. Its effect is readily described after
expanding $Q^{(K)}$ and $Q^{(K)}_\psi$ in the basis (\ref{Nbasis}):
\begin{align}
Q^{(K)}&=2\,m\,H_0=\frac{m}{2}\,(\mathcal{N}_{(1,0)}+\mathcal{N}_{(-1,0)}+\mathcal{N}_{(0,1)}+\mathcal{N}_{(0,-1)})\,,\label{KNNUT2}\\
Q^{(K)}_\psi&=2\,\alpha \,m\,\mathbb{K}_\bullet=\frac{\alpha
m}{2}\,\left(\mathcal{N}_{\left(\frac{1}{2},\,\frac{3}{2}\right)}+\mathcal{N}_{\left(\frac{1}{2},\,-\frac{1}{2}\right)}
+\mathcal{N}_{\left(-\frac{1}{2},\,-\frac{3}{2}\right)}+\mathcal{N}_{\left(-
\frac{1}{2},\,\frac{1}{2}\right)}\right)
\end{align}
If we have a solution whose Noether-charge matrix expands on the
generators $\mathcal{N}_{(\pm 1,0)},\,\mathcal{N}_{(0,\pm 1)}$, its
compact component will expand in the generators
$\mathbb{K}_0,\,\mathbb{K}^1$ only, see Eq. (\ref{q0p1}), and thus,
according to (\ref{KNNUT}), will only have charges $q_0,\,p^1$. The values of these charges can be read off from the general expression in Eq. \ref{KNNUT}. The Noether-matrix $Q^{(K)}$ in
(\ref{KNNUT2}) defines a (electrically and magnetically) neutral solution since nilpotent generators
with opposite gradings are summed and thus their compact components
cancel against one another.

 Upon acting on the solution by means of $\mathcal{O}_{(q_0,\,p^1)}$, and  using  Eq.
(\ref{abaction}), we find the transformed matrices $Q,\,Q_\psi$:
\begin{align}
Q&=\mathcal{O}_{(q_0,\,p^1)}^{-1}Q^{(K)}
\mathcal{O}_{(q_0,\,p^1)}=\frac{m}{2}\,\left(\frac{1}{\beta_0}\mathcal{N}_{(1,0)}+\beta_0\,
\mathcal{N}_{(-1,0)}+\frac{1}{\beta_1}\mathcal{N}_{(0,1)}+\beta_1\,\mathcal{N}_{(0,-1)}\right)\,,
\label{KNNUT3}\\
Q_\psi&=\mathcal{O}_{(q_0,\,p^1)}^{-1}Q^{(K)}_\psi
\mathcal{O}_{(q_0,\,p^1)}\nonumber\\
&=\frac{\alpha
m}{2}\,\left(\frac{1}{\beta_0^{\frac{1}{2}}\,\beta_1^{\frac{3}{2}}}\,\mathcal{N}_{\left(\frac{1}{2},\,\frac{3}{2}\right)}+
\sqrt{\frac{\beta_1}{\beta_0}}\,\mathcal{N}_{\left(\frac{1}{2},\,-\frac{1}{2}\right)}
+\beta_0^{\frac{1}{2}}\,\beta_1^{\frac{3}{2}}\,\mathcal{N}_{\left(-\frac{1}{2},\,-\frac{3}{2}\right)}+\sqrt{\frac{\beta_0}{\beta_1}}\,\mathcal{N}_{\left(-
\frac{1}{2},\,\frac{1}{2}\right)}\right),\,\label{KNNUT3psi}
\end{align}
from which we can infer the physical properties of the transformed
solution by projecting  (\ref{KNNUT3}) and (\ref{KNNUT3psi}) on the relevant generators. In
particular we find:
\begin{align}
M_{ADM}&=\frac{m}{8}\,\left( \frac{1}{{\beta_0}} + {\beta_0} +
\frac{3}{{\beta_1}} + 3\,{\beta_1}
\right)\,,\,\,p^0=0\,,\,\,p^1=m\frac{\left( -1 + {{\beta_1}}^2
\right) }{2\,{\sqrt{2}}\,{\beta_1}}\,\,,\,\,\,q_0=m\,\frac{1 -
{{\beta_0}}^2}{2\,{\sqrt{2}}\,{{\beta}_0}}\,,\,\,q_1=0\,,\nonumber\\
\Sigma_1&=0\,,\,\,\,\Sigma_2={\sqrt{3}}\,m\,\frac{ (\beta_0-\beta_1)(1-\beta_0\beta_1) }{8\,{\beta_0}\,{{\beta}_1}}\,\,,\,\,\,\ell=0\,\,,\,\,\,M_\varphi=\alpha\,m\,\frac{1 + 3\,{{\beta_1}}^2 +
      {\beta_0}\,{\beta_1}\,\left( 3 + {{\beta_1}}^2 \right)   }{8\,{\sqrt{{\beta_0}}}\,
    {{\beta_1}}^{\frac{3}{2}}}\,.
\end{align}
Before writing the explicit solution, let us analyze the three
relevant singular limits from the point of view of the physical
properties at infinity.
To make contact with the discussion of Sect. \ref{section2}, we recall that in the $T^3$-model there are just two regular orbits of the electric-magnetic representation ${\bf R}={\bf \frac{3}{2}}$ of $G_4={\rm SL}(2,\mathbb{R})$:
\begin{align}
\mbox{BPS}&:\,\,I_4(p,q)>0\,,\nonumber\\
\mbox{non-BPS}&:\,\,I_4(p,q)<0\,.\nonumber
\end{align}
where the general form of $I_4$ is (see also \cite{Ferrara:2010ug}):
\begin{eqnarray}
 I_4(p,q) & = &-(q_0p^0)^2-2\,q_0p^0q_1p^1+\frac{1}{3}\,(q_1p^1)^2+4q_0(p^1)^3-\frac{4}{27}\,p^0(q_1)^3\,.\label{quarticinvt3}
\end{eqnarray}

\subsection{Limits}\label{limitsec}
\paragraph{Non-BPS under-rotating.} This corresponds to taking:
\begin{align}
\beta_\ell=m\,\alpha_\ell\,\,,\,\,\,\alpha=m\,\Omega\,,\label{nbpspre}
\end{align}
and then sending $m\rightarrow 0$ while keeping
$\alpha_\ell,\,\Omega$ fixed. In this limit the Harrison
transformation, as well as the original Kerr solution,  become
singular and the semisimple generators $Q,\,Q_\psi$ become
nilpotent. To appreciate the details of this procedure, let us
rewrite $Q,\,Q_\psi$ in terms of $\alpha_\ell,\,\Omega$:
\begin{align}
Q&=\frac{1}{2}\left(\frac{1}{\alpha_0}\mathcal{N}_{(1,0)}+m^2\,\alpha_0\,\mathcal{N}_{(-1,0)}+\frac{1}{\alpha_1}\mathcal{N}_{(0,1)}+m^2\,\alpha_1\,\mathcal{N}_{(0,-1)}\right)\,,
\label{KNNUT4}\\
Q_\psi&=\frac{\Omega}{2}\,\left(\frac{1}{\alpha_0^{\frac{1}{2}}\,\alpha_1^{\frac{3}{2}}}\,
\mathcal{N}_{\left(\frac{1}{2},\,\frac{3}{2}\right)}+
m^2\,\sqrt{\frac{\alpha_1}{\alpha_0}}\,\mathcal{N}_{\left(\frac{1}{2},\,-\frac{1}{2}\right)}
+m^4\,\alpha_0^{\frac{1}{2}}\,\alpha_1^{\frac{3}{2}}\,\mathcal{N}_{\left(-\frac{1}{2},\,-\frac{3}{2}\right)}+m^2\,\sqrt{\frac{\alpha_0}{\alpha_1}}\,\mathcal{N}_{\left(-
\frac{1}{2},\,\frac{1}{2}\right)}\right),\,
\end{align}
In the $m\rightarrow 0$ limit we find:
\begin{equation}
Q^{(0)}=\lim_{m\rightarrow
0}Q=\frac{1}{2}\left(\frac{1}{\alpha_0}\mathcal{N}_{(1,0)}+\frac{1}{\alpha_1}\mathcal{N}_{(0,1)}\right)\,\,\,\,;\,\,\,\,\,\,Q_\psi^{(0)}=\lim_{m\rightarrow
0}Q_\psi=\frac{\Omega}{2\,\alpha_0^{\frac{1}{2}}\,\alpha_1^{\frac{3}{2}}}\,
\mathcal{N}_{\left(\frac{1}{2},\,\frac{3}{2}\right)}\,.
\end{equation}
Notice that $Q^{(0)}$ is a step-3 nilpotent generator
($Q^{(0)\,3}=0$), while $Q_\psi^{(0)}$ is step-2. The resulting
solution is extremal for the reasons explained in the previous Section.
In the above limit the physical quantities stay finite and read:
\begin{align}
M_{ADM}&=\frac{1}{8}\,\left( \frac{1}{{\alpha_0}}  +
\frac{3}{{\alpha_1}}\right)\,,\,\,p^0=0\,,\,\,p^1=-\frac{1
}{2\,{\sqrt{2}}\,{\alpha_1}}\,\,,\,\,\,q_0=\frac{1}{2\,{\sqrt{2}}\,{\alpha_0}}\,,\,\,q_1=0\,,\nonumber\\
\Sigma_1&=0\,,\,\,\,\Sigma_2=- \frac{{\sqrt{3}}}{8}\,\frac{
{\alpha_1} -
        {\alpha_0}  }{{\alpha_0}\,{\alpha_1}}\,\,,\,\,\,\ell=0\,\,,\,\,\,M_\varphi=\frac{\Omega  }{8\,{\sqrt{{\alpha_0}}}\,
    {{\alpha_1}}^{\frac{3}{2}}}=\frac{\Omega}{2}\,\sqrt{|I_4(p,q)|}\,.\label{nbpsch}
\end{align}
Notice that the quartic invariant
$I_4(p^\Lambda,q_\Lambda)=4\,q_0\,(p^1)^3$   is negative and the angular momentum is different
from zero. This solution is  the (single-center) under-rotating
almost-BPS solution studied in \cite{Goldstein:2008fq,Bena:2009ev}.
The limiting procedure defined above generalizes therefore the one
studied by Rasheed and Larsen in the (dilaton-Maxwell) $D=4$ Kaluza-Klein
theory and, as we shall see, generalizes to all symmetric cubic
supergravity models.
\paragraph{Static BPS.}
Consider now the rescaling:
\begin{align}
\beta_0&=m\,\alpha_0\,\,,\,\,\,\beta_1=\alpha_1/m\,\,,\,\,\,\alpha=m\,\Omega\,.
\end{align}
The matrices $Q,\,Q_\psi$ now read:
\begin{align}
Q&=\frac{1}{2}\left(\frac{1}{\alpha_0}\mathcal{N}_{(1,0)}+m^2\,\alpha_0\,\mathcal{N}_{(-1,0)}+\frac{m^2}{\alpha_1}\mathcal{N}_{(0,1)}+\alpha_1\,\mathcal{N}_{(0,-1)}\right)\,,
\label{KNNUT4}\\
Q_\psi&=\frac{\Omega}{2}\,\left(\frac{m^3}{\alpha_0^{\frac{1}{2}}\,\alpha_1^{\frac{3}{2}}}\,
\mathcal{N}_{\left(\frac{1}{2},\,\frac{3}{2}\right)}+
m\,\sqrt{\frac{\alpha_1}{\alpha_0}}\,\mathcal{N}_{\left(\frac{1}{2},\,-\frac{1}{2}\right)}
+m\,\alpha_0^{\frac{1}{2}}\,\alpha_1^{\frac{3}{2}}\,\mathcal{N}_{\left(-\frac{1}{2},\,-\frac{3}{2}\right)}+m^3\,\sqrt{\frac{\alpha_0}{\alpha_1}}\,\mathcal{N}_{\left(-
\frac{1}{2},\,\frac{1}{2}\right)}\right),\,
\end{align}
In the limit $m\rightarrow 0$ we find the following matrices:
\begin{equation}
Q^{(0)}=\lim_{m\rightarrow
0}Q=\frac{1}{2}\left(\frac{1}{\alpha_0}\mathcal{N}_{(1,0)}+\frac{1}{\alpha_1}\mathcal{N}_{(0,-1)}\right)\,\,\,\,;\,\,\,\,\,\,Q_\psi^{(0)}=\lim_{m\rightarrow
0}Q_\psi=0\,.
\end{equation}
Since $Q^{(0)}$ is now a step-3 nilpotent generator but
$Q_\psi^{(0)}=0$ the solution is still extremal, though static. The
related  physical quantities are readily computed:
\begin{align}
M_{ADM}&=\frac{1}{8}\,\left( \frac{1}{{\alpha_0}}  +
3\,\alpha_1\right)\,,\,\,p^0=0\,,\,\,p^1=\frac{{\alpha_1}
}{2\,{\sqrt{2}}}\,\,,\,\,\,q_0=\frac{1}{2\,{\sqrt{2}}\,{\alpha_0}}\,,\,\,q_1=0\,,\nonumber\\
\Sigma_1&=0\,,\,\,\,\Sigma_2=- \frac{{\sqrt{3}}}{8}\,\frac{ {1-
        {\alpha_0}\alpha_1}   }{{\alpha_0}}\,\,,\,\,\,\ell=0\,\,,\,\,\,M_\varphi=0\,.
\end{align}
We see that the quartic invariant
$I_4(p^\Lambda,q_\Lambda)=4\,q_0\,(p^1)^3$ is now positive and the
solution is the single center BPS one.
\paragraph{Rotating BPS.}
This limit is obtained by first defining
\begin{align}
\beta_0&=m\,\alpha_0\,\,,\,\,\,\beta_1=\alpha_1/m\,\,,\,\,\,
%\alpha=\Omega\,.
\end{align}
and then  sending $m\rightarrow 0$. As opposed to the previous case,
$\alpha$ is kept fixed. As a consequence of this, $Q_\psi^{(0)}$ is a
non-vanishing nilpotent matrix:
\begin{equation}
Q_\psi^{(0)}=\lim_{m\rightarrow 0}Q_\psi=\frac{\alpha}{2}\,
\left(\sqrt{\frac{\alpha_1}{\alpha_0}}\,\mathcal{N}_{\left(\frac{1}{2},\,-\frac{1}{2}\right)}+
\alpha_0^{\frac{1}{2}}\,\alpha_1^{\frac{3}{2}}\,\mathcal{N}_{\left(-\frac{1}{2},\,-\frac{3}{2}\right)}\right)\,,
\end{equation}
while $Q^{(0)}$ is the same as in the previous case. Now both
$Q^{(0)},\,Q^{(0)}_\psi$ are step-3 nilpotent. The regularity condition  (\ref{regcond}), in the $m\to 0$ limit,  is not satisfied since the left hand side vanishes while the right hand one remains finite, implying that the solution is singular.\par
The physical quantities associated with this solution are the same
as in the previous limit, but for the angular momentum which now
reads:
\begin{align}
M_\varphi=\frac{\alpha \,{\sqrt{{\alpha_1}}}\,\left( 3 + {\alpha_0}\,{\alpha_1} \right) }{8\,{\sqrt{{\alpha_0}}}}\,.\label{mphibpsr}
\end{align}
This is a rotating BPS solution of the kind first studied in
\cite{Behrndt:1997ny}.
\subsection{The Solution}\label{solution}
In this subsection we give the explicit $D=4$ non-extremal axion-dilaton solution corresponding to the two normal sets of charges. We start from the non-extremal Kerr solution, given in Appendix \ref{Afin}, whose $D=3$ description is given in terms of the following scalar fields:
\begin{align}
\Phi^I_{(K)}(r,\theta)&:\,\,\,e^{2U}=\frac{\tilde{\Delta}}{\rho^2}\,,\,\,\, a=\frac{\alpha m}{\rho^2}\,\cos\theta \,,\,\,\, \mathcal{Z}^M=0 \,,
\end{align}
and we apply to it the transformations $\mathcal{O}_{(q_0,\,p^1)}$ and $\mathcal{O}_{(p^0,\,q_1)}$.
The $D=4$ stationary metric has the general form (\ref{statmet}) with the three dimensional metric given by
\begin{equation}
(g^{(3)}_{ij})=\left(\begin{matrix}\frac{\tilde{\Delta}}{\Delta}& 0 & 0\cr 0 & \tilde{\Delta} & 0\cr 0 & 0 & \Delta\,\sin^2\theta \end{matrix}\right)\,,
\end{equation}
where, as usual (see Appendix \ref{Afin}), we define the quantities:
\begin{equation}
\tilde{\Delta}=r(r-2 m)+\alpha^2\,\cos^2\theta \,\,;\,\,\,\,\Delta=r(r-2 m)+\alpha^2\,\,;\,\,\,\,\rho^2=r^2 +\alpha^2\,\cos^2\theta \,.
\end{equation}
\subsubsection{The $q_0,\,p^1$-Case}
Here we give and discuss the non-extremal, rotating axion-dilaton solution generated by applying the Harrison transformation $\mathcal{O}_{(q_0,\,p^1)}$ to the Kerr solution.
The $D=3$ scalars $\Phi^I(r,\theta)$ describing the transformed solution are obtained in terms of $\Phi^I_{(K)}(r,\theta)$ by solving the matrix equation:
\begin{equation}
\mathcal{M}[\Phi^I(r,\theta)]=\mathcal{O}_{(q_0,\,p^1)}\mathcal{M}[\Phi^I_{(K)}(r,\theta)]\mathcal{O}_{(q_0,\,p^1)}^T\,.\label{mateqq0p1}
\end{equation}
It is convenient, in order to write  $\Phi^I(r,\theta)$, to introduce the following new quantities:
\begin{equation}
\eta_\ell=r-m+\frac{m}{2}\,(\beta_\ell+\beta_\ell^{-1})\,.
\end{equation}
We then find, by solving Eq. (\ref{mateqq0p1}):
\begin{align}
e^{-4U}&=\frac{\tilde{\rho}^4}{\tilde{\Delta}^2}\,,\nonumber\\
\mathcal{Z}^0&=\frac{e^{4U}\,m}{2\sqrt{2}\tilde{\Delta}^2\,\beta_0}\,(\beta_0^2-1)\,\left(\eta_1\,
(\eta_1^2+\alpha^2\cos^2\theta )-
\frac{m \alpha ^2 \cos ^2\theta  \left(\beta_1
^2-1\right)^2 \left(\beta_0^2 \beta_1^2-1\right)}{8 \left(\beta_0^2-1\right) \beta_1^3}\right)\,,\nonumber\\
\mathcal{Z}^1&=\frac{e^{4U}}{\tilde{\Delta}^2}\frac{m \alpha  \cos \theta  \left(\beta_0 \beta_1+1\right) \left(\beta_1^2-1\right)}{4 \sqrt{2}
   \sqrt{\beta_0} \beta_1^{3/2}}\,\left(\eta_1^2+\alpha^2\cos^2\theta -\frac{m \left(\beta_0 \beta_1-1\right) \left(\beta_1^2-1\right)}{2 \beta_1 \left(\beta_0 \beta_1+1\right)}\,\eta_1\right)\,,\nonumber\\
   \mathcal{Z}_0&=-\frac{e^{4U}}{\tilde{\Delta}^2}\frac{m \alpha  \cos \theta  \, K^{(+)}_-}{4 \sqrt{2} \sqrt{\beta_0} \beta_1^{3/2}}\,\left(\eta_1\eta_0+\alpha^2\cos^2\theta -\frac{m \left(\beta_0 \beta_1-1\right) \left(\beta_1^2-1\right)^2}{2 \beta_1\, K^{(+)}_-}\,\eta_0\right)\,,\nonumber\\
     \mathcal{Z}_1&=\frac{3 e^{4U}\,m}{2\sqrt{2}\tilde{\Delta}^2\,\beta_1}\,(\beta_1^2-1)\,\left(\eta_1\,
     (\eta_1\eta_0+\alpha^2\cos^2\theta )-\frac{m \alpha ^2 \cos ^2\theta  \left(\beta_1^2-1\right) \left(\beta_0^2 \beta_1^2-1\right)}{8 \beta_0 \beta_1^2}\right)\,,\nonumber\\
   e^\varphi&=\frac{\tilde{\Delta}\,e^{-2U}}{\eta_1^2+\alpha^2\cos^2\theta }\,\,,\,\,\,
   y=-\frac{m \alpha  \cos \theta  \left(\beta_0 \beta
   _2-1\right) \left(\beta_1^2-1\right)}{4
   \sqrt{\beta_0} \beta_1^{3/2}\,(\eta_1^2+\alpha^2\cos^2\theta )}\,,\nonumber\\
   a&=\frac{e^{4U}}{\tilde{\Delta}^2}\frac{m \alpha  \cos \theta  \, K^{(+)}_+}{4\,\sqrt{\beta_0} \beta_1^{3/2}}\,\left(\frac{\eta_1(\eta_0+\eta_1)}{2}+\alpha^2\cos^2\theta -\frac{m^2 \left(\beta_0+\beta_1\right) \left(\beta_0 \beta_1-1\right)^2 \left(\beta_1^2-1\right)^2}{8 \beta_0 \beta_1^2\,K^{(+)}_+}\right)\,.
\end{align}
where we have defined:
\begin{align}
\tilde{\rho}^4&=(\eta_1^2+\alpha^2\cos^2\theta )(\eta_1\eta_0+\alpha^2\cos^2\theta )-\frac{m^2 \alpha^2\cos^2\theta \, \left(\beta_0 \beta_1-1\right)^2
   \left(\beta_1^2-1\right)^2}{16 \beta_0 \beta_1^3}\,,\nonumber\\
   K^{(+)}_\pm&=\pm (1+ 3 \beta_1^2)+\beta_0 \, \beta_1\left(\beta_1^2+3\right)=\frac 12\left[(\beta_0\pm 1)(\beta_1+1)^3 + (\beta_0\mp 1)(\beta_1-1)^3\right]\,.
\end{align}
To derive the $D=4$ metric and vector fields we use the dualization formulae:
\begin{align}
F^0&=d\omega=-e^{-4U}\,*_3(da+\mathcal{Z}^T\mathbb{C}\,d\mathcal{Z})\,,\nonumber\\
\mathbb{F}&=\left(\begin{matrix}F^\Lambda \cr G_\Lambda\end{matrix}\right)=e^{-2U}\,\mathbb{C}\,\mathcal{M}_4\,*_3d\mathcal{Z}+d\mathcal{Z}\wedge (dt+\omega)\,,\label{dual}
\end{align}
where $*_3$ is the Hodge-duality operation with respect to the $D=3$ metric $g^{(3)}$.
We find:
\begin{equation}
\omega=\frac{m \alpha  \sin ^2\theta }{4 \sqrt{\beta_0}
   \beta_1^{3/2}\,\tilde{\Delta}}\left(K^{(+)}_+\,r-m \left(\beta_0-1\right) \left(\beta_1-1\right)^3\right)=\frac{\sin^2\theta }{\tilde{\Delta}}\,\left(2 M_\varphi  r\, -\frac{m^2 \alpha  \left(\beta_0-1\right) \left(\beta_1-1\right)^3}{4
   \sqrt{\beta_0} \beta_1^{3/2}}\right)\,.
\end{equation}
We can also \emph{locally} integrate $\mathbb{F}^M$ to a symplectic vector of electric and magnetic potentials $A^M_\mu$:
\begin{equation}
\mathbb{F}^M=dA^M\,\,;\,\,\,\,A^M=\mathcal{Z}^M\,(dt+\omega)+A^M_i\,dx^i\,,
\end{equation}
where $A^M_i\,dx^i=A^M_\varphi\,d\varphi$ are solutions to the differential equations (we suppress the symplectic index $M$ for the sake of notational simplicity):
\begin{align}
\partial_r A_\varphi&=-e_3\,e^{-2U}\,\mathbb{C}\,\mathcal{M}_4\,\partial^\theta\mathcal{Z}- \mathcal{Z}\,\partial_r \omega_\varphi\,,\nonumber\\
\partial_\theta A_\varphi&=e_3\,e^{-2U}\,\mathbb{C}\,\mathcal{M}_4\,\partial^r\mathcal{Z}- \mathcal{Z}\,\partial_\theta\omega_\varphi\,,
\end{align}
which directly follow from (\ref{dual}) (indices are raised and lowered using $g^{(3)}$).
The integration of the above equations yields:
\begin{align}
A^0_\varphi&=\frac{m \alpha  \sin ^2\theta } {4 \sqrt{2} \tilde{\Delta} \sqrt{\beta_0} \beta_1^{3/2}}\left(m \left(\beta_0+1\right) \left(\beta_1-1\right)^3-r K^{(+)}_-\right)\,,\nonumber\\
 A^1_\varphi&=-\frac{m \, \cos \theta  \left(\beta_1^2-1\right)}{2 \sqrt{2} \beta_1\,}\,\frac{\Delta }{\tilde{\Delta}}\,\,;\,\,\,
A_{0\,\varphi}=\frac{m \, \cos \theta  \left(\beta_0^2-1\right)}{2 \sqrt{2} \beta_0\,}\,\frac{\Delta }{\tilde{\Delta}}\,,\nonumber\\
   A_{1\,\varphi}&=\frac{3 m \alpha  \sin ^2\theta  \left(\beta_1^2-1\right) \left(m \left(\beta_0-1\right)
   \left(\beta_1-1\right)-r \left(\beta_0 \beta_1+1\right)\right)}{4 \sqrt{2} \sqrt{\beta_0}
   \beta_1^{3/2}\,\tilde{\Delta}}\,.
\end{align}
\subsubsection{Extremal Limits}
We now apply to the above solution the limit-prescriptions given in the first part of  Sect.\ref{limitsec}.
It is useful to rewrite, in the $m\rightarrow 0$ limit, the $\eta_\ell$ in terms of harmonic functions:
\begin{equation}
\eta_\ell=r\,H_\ell\,\,;\,\,\,\,H_0=1+\sqrt{2}\,\frac{|q_0|}{r}\,\,;\,\,\,H^1=1+\sqrt{2}\,\frac{|p^1|}{r}\,,
\end{equation}
In the same limit we also have $\tilde{\Delta}\rightarrow r^2$.
\paragraph{non-BPS under-rotating.}
Using the redefinitions (\ref{nbpspre}) and the identifications (\ref{nbpsch}) in the $m\rightarrow 0$,  we find:
\begin{align}
e^{-4U}&=H_0\,(H^1)^3-4\frac{M_\varphi^2\,\cos^2\theta }{r^4}\,\,;\,\,\,T=-\frac{1}{(H^1)^2}\left(2\,
\frac{M_\varphi\,\cos\theta }{r^2}+{\rm i}\,e^{-2U}\right)\,,\nonumber\\
%a&=2\,e^{4U}\,
%\frac{M_\varphi\,\cos\theta }{r^2}\,\left(\frac{3H^1+H_0}{2}-1\right)\,\,\,;\,\,\,
\omega&=\frac{2\,M_\varphi}{r}
\sin^2\theta \,d\varphi\,,\nonumber\\
A^0&=-\frac{e^{4U}\,q_0}{r}\,\left((H^1)^3-\frac{2\sqrt{2}\, M_\varphi^2}{q_0\,r^3}\,\cos^2\theta \right)(dt+\omega)+\frac{\sqrt{2}\,M_\varphi}{r}\,\sin^2\theta \,d\varphi\,,\nonumber\\
A^1&=-e^{4U}\,\frac{\sqrt{2}\,M_\varphi}{r^2}\,\cos\theta H^1\,(dt+\omega)-p^1\,\cos\theta \,d\varphi\,,\nonumber\\
A_0&=e^{4U}\,\frac{\sqrt{2}\,M_\varphi}{r^2}\,\cos\theta H_0\,(dt+\omega)-q_0\,\cos\theta \,d\varphi\,,\nonumber\\
A_1&=3\frac{e^{4U}\,p^1}{r}\,\left(H_0\,(H^1)^2+\frac{2\sqrt{2}\, M_\varphi^2}{p^1\,r^3}\,\cos^2\theta \right)(dt+\omega)+3\,\frac{\sqrt{2}\,M_\varphi}{r}\,\sin^2\theta \,d\varphi\,,
\end{align}
which fits the general form given in \cite{Bena:2009ev} for the minimal set of parameters modulo $G_{(3)}$-action.
\paragraph{Rotating-BPS.} Below we write the full rotating-BPS (singular) solution.
\begin{align}
e^{-4U}&=\frac{\tilde{\rho}^4}{\tilde{\Delta}^2}\,,\nonumber\\
\tilde{\rho}^4&=r^4\,\left([H_0 H^1+\frac{\alpha^2}{r^2}\,\cos^2\theta ][ (H^1)^2+\frac{\alpha^2}{r^2}\,\cos^2\theta ]-\frac{\alpha^2\,\cos^2\theta }{2 r^4}\,(p^1-q_0)^2\frac{p^1}{q_0}\right)\,,\nonumber\\
\tilde{\Delta}^2&=r^2+\alpha^2\,\cos^2\theta \,,\nonumber\\
\omega&=\frac{ \sin^2\theta}{\tilde{\Delta}}\,(2\,M_\varphi r+{\alpha}\,\sqrt{I_4(p,q)})\,.\nonumber\\
T&=-\frac{1}{r^2 (H^1)^2+{\alpha^2}\,\cos^2\theta }\,\left(\frac{\alpha \cos\theta (p^1-q_0)}{\sqrt{2} }\sqrt{\frac{p^1}{q_0}}+{\rm i}\,e^{-2U}\,\tilde{\Delta}\right)\,,\nonumber\\
A^0&=-\frac{q_0\,r^3}{\tilde{\rho}^4}\,\left(H^1[(H^1)^2+\frac{\alpha^2}{r^2}\,\cos^2\theta ]+\frac{\alpha^2\cos^2\theta  p^1}{2\sqrt{2} r^3}\left(\left(\frac{p^1}{q_0}\right)^2-1\right)\right)\,(dt+\omega)+\nonumber\\
&+\frac{\alpha\sin^2\theta }{2\tilde{\Delta}}\,\sqrt{q_0p^1}\left(2\sqrt{2}\,p^1-r\left(\frac{p^1}{q_0}-3\right)\right)\,d\varphi\,,\nonumber\\
A^1&=\frac{r^2\alpha\cos\theta }{2\tilde{\rho}^4}\,\sqrt{q_0p^1}\left(\frac{p^1}{q_0}+1\right)\left((H^1)^2+\frac{\alpha^2}{r^2}\,\cos^2\theta
-\frac{\sqrt{2}\,p^1\,H^1}{r}\frac{p^1-q_0}{p^1+q_0}\right)\,(dt+\omega)+\nonumber\\
&-\frac{\Delta }{\tilde{\Delta}}\,p^1\cos\theta \,d\varphi\,,\nonumber\\
A_0&=-\frac{r^2\alpha\cos\theta }{2\tilde{\rho}^4}\,\sqrt{q_0p^1}\left(\frac{p^1}{q_0}-3\right)\left(
H^1 H_0+\frac{\alpha^2}{r^2}\,\cos^2\theta
-\frac{2 \sqrt{2}\,p^1\,H_0}{r}\frac{p^1-q_0}{p^1-3q_0}\right)\,(dt+\omega)+\nonumber\\
&-\frac{\Delta }{\tilde{\Delta}}\,q_0\cos\theta \,d\varphi\,,\nonumber\\
A_1&=3\frac{p^0\,r^3}{\tilde{\rho}^4}\,\left(H^1[H^1 H_0+\frac{\alpha^2}{r^2}\,\cos^2\theta ]-\frac{\alpha^2\cos^2\theta  q_0}{2\sqrt{2} r^3}\left(\left(\frac{p^1}{q_0}\right)^2-1\right)\right)\,(dt+\omega)+\nonumber\\
&-3\frac{\alpha\sin^2\theta }{2\tilde{\Delta}}\,\sqrt{q_0p^1}\,\left(2\sqrt{2}\,p^1+
r\left(\frac{p^1}{q_0}+1\right)\right)\,d\varphi\,,
\end{align}
where $M_\varphi$ is given in Eq.s (\ref{mphibpsr}) while the quartic invariant $I_4(p,q)$, given in Eq. (\ref{quarticinvt3}), for this choice of the charges  reads: $I_4(p,q)=4q_0(p^1)^3$.
Setting $\alpha\rightarrow 0$ we recover the known (regular) static BPS solution \cite{Behrndt:1997ny}.

\subsubsection{The $p^0,\,q_1$-Case}
Let us now transform the Kerr solution using the transformation $\mathcal{O}_{(p^0,q_1)}$ generated by $\mathcal{J}'_\ell=\{J_{p^0},\,J_{q_1}\}$:
\begin{equation}
\mathcal{O}_{(p^0,q_1)}=e^{\log(\beta_\ell)\,\mathcal{J}'_\ell}\,.
\end{equation}
The computation of the charges at infinity proceeds along the same lines as in the $q_0,\,p^1$ case.
We find:
\begin{align}
M_{ADM}&=\frac{1}{8} m \left(\beta_0+3 \beta_1+\frac{3}{\beta_1}+\frac{1}{\beta_0}\right)\,,\,\,p^0=\frac{m \left(\beta_0^2-1\right)}{2 \sqrt{2} \beta_0}\,,\,\,p^1=0\,\,,\,\,\,q_0=0\,,\,\,q_1=-\frac{3 m \left(\beta_1^2-1\right)}{2 \sqrt{2} \beta_1}\,,\nonumber\\
\Sigma_1&=0\,,\,\,\,\Sigma_2=\frac{\sqrt{3} m (\beta_0 -\beta_1)(\beta_0\beta_1 - 1)}{8 \beta_0 \beta_1}\,\,,\,\,\,\ell=0\,\,,\,\,\,M_\varphi=\frac{m \alpha \, K^{(-)}_+}{8 \sqrt{\beta_0} \beta_1^{3/2}}\,.
\end{align}
where we found useful to introduce the combinations of charges
\begin{eqnarray}
K^{(-)}_\pm&=
\beta_0 + 3 \beta _0
   \beta _1^2\pm\left(\beta
   _1^3+3\beta_1\right)= \frac 12\left[(\beta_0\pm 1)(\beta_1+1)^3 - (\beta_0\mp 1)(\beta_1-1)^3\right]
   \,.
   \end{eqnarray}
By solving the matrix equation:
  \begin{equation}
\mathcal{M}[\Phi^I(r,\theta)]=\mathcal{O}_{(p^0,q_1)}\mathcal{M}[\Phi^I_{(K)}(r,\theta)]\mathcal{O}_{(p^0,q_1)}^T\,,\label{mateqp0q1}
\end{equation}
we derive the $D=3$ scalar fields $\Phi^I(r,\theta)$:
\begin{align}
e^{-4U}&=\frac{\tilde{\rho}^{\prime 4}}{\tilde{\Delta}^2}\,,\nonumber\\
\mathcal{Z}^0&=\frac{e^{4U}}{\tilde{\Delta}^2}\frac{m \alpha  \cos \theta  \,K^{(-)}_-}{4 \sqrt{2} \sqrt{\beta_0}
   \beta_1^{3/2}}\,\left(\eta_1\eta_0+\alpha^2\cos^2\theta +\frac{m \left(\beta_1-\beta_0\right) \left(\beta_1^2-1\right)^2}{2 \beta_1 \,K^{(-)}_-}\,\eta_0\right)\,,\nonumber\\
\mathcal{Z}^1&=\frac{e^{4U}}{\tilde{\Delta}^2}\,\frac{m \left(\beta_1^2-1\right)}{2 \sqrt{2} \beta_1}\,\left(\eta_1\,
     (\eta_1\eta_0+\alpha^2\cos^2\theta )+\frac{m \alpha ^2 \cos ^2\theta  \left(\beta_0^2-\beta_1^2\right) \left(\beta_1^2-1\right)}{8
   \beta_0 \beta_1^2}\right)\,,\nonumber\\
   \mathcal{Z}_0&=\frac{e^{4U}}{\tilde{\Delta}^2}\,\frac{m \left(\beta_0^2-1\right)}{2 \sqrt{2} \beta_0}\,\left(\eta_1\,
     (\eta_1^2+\alpha^2\cos^2\theta )+\frac{m \alpha ^2 \cos ^2\theta  \left(\beta_1^2-1\right)^2 \left(\beta_1^2-\beta_0^2\right)}{8 \left(\beta_0^2-1\right) \beta_1^3}\right)\,,\nonumber\\
     \mathcal{Z}_1&=-\frac{e^{4U}}{\tilde{\Delta}^2}\,\frac{3 m \alpha  \cos \theta  \left(\beta_0+\beta_1\right) \left(\beta_1^2-1\right)}{4 \sqrt{2}
   \sqrt{\beta_0} \beta_1^{3/2}}\,\left(\eta_1^2+\alpha^2\cos^2\theta +\frac{m \left(\beta_0-\beta_1\right) \left(\beta_1^2-1\right)}{2 \beta_1 \left(\beta_0+\beta_1\right)}\,\eta_1\right)\,,\nonumber\\
   e^\varphi&=\frac{\tilde{\Delta}\,e^{-2U}}{\eta_1\eta_0+\alpha^2\cos^2\theta }\,\,,\,\,\,
   y=\frac{m \alpha  \cos \theta  \left(\beta_1-\beta_0\right) \left(\beta_1^2-1\right)}{4 \sqrt{\beta_0} \beta_1^{3/2}(\eta_1\eta_0+\alpha^2\cos^2\theta )}\,,\nonumber\\
   a&=\frac{e^{4U}}{\tilde{\Delta}^2}\frac{m \alpha  \cos \theta  K^{(-)}_+}{4
   \sqrt{\beta_0
   } \beta_1^{3/2}}\,\left(\frac{\eta_1(\eta_0+\eta_1)}{2}+\alpha^2\cos^2\theta -\frac{m^2 \left(\beta_0-\beta_1\right)^2
   \left(\beta_0 \beta_1+1\right) \left(\beta_1^2-1\right)^2}{8 \beta_0 \beta_1^2 K^{(-)}_+}\right)\,.
\end{align}
where:
\begin{align}
\tilde{\rho}^{\prime 4}&=\left(\eta_1\eta_0+\alpha^2\cos^2\theta \right)\left(\eta_1^2+\alpha^2\cos^2\theta
-y^2\,(\eta_1\eta_0+\alpha^2\cos^2\theta )\right)\,.
\end{align}
Integrating $a,\,\mathcal{Z}^M$ to the $D=4$ metric component and vector  fields we find:
\begin{align}
\omega&=\frac{\sin^2\theta }{\tilde{\Delta}}\,\left(2 M_\varphi  r\, +\frac{m^2 \alpha  \left(\beta_0-1\right) \left(\beta_1-1\right)^3}{4
   \sqrt{\beta_0} \beta_1^{3/2}}\right)\,\,;\,\,\,
   A^M=\mathcal{Z}^M\,(dt+\omega)+A^M_\varphi\,d\varphi\,,
\end{align}
where:
\begin{align}
A_\varphi^0&=-\frac{\Delta}{\tilde{\Delta}}\,\frac{m \left(\beta_0^2-1\right)}{2 \sqrt{2} \beta_0}\,\cos\theta \,\,,\,\,A_\varphi^1=-\frac{m \alpha  \sin ^2\theta  \left(m \left(\beta_0-1\right) \left(\beta_1-1\right)+r
   \left(\beta_0+\beta_1\right)\right) \left(\beta_1^2-1\right)}{4 \sqrt{2} \tilde{\Delta}
   \sqrt{\beta_0} \beta_1^{3/2}}\,,\nonumber\\
   A_{0\varphi}&=-\frac{m \alpha  \sin ^2\theta  \left(m \left(\beta_0+1\right) \left(\beta_1-1\right)^3+r K^{(-)}_-\right)}{4 \sqrt{2}
   \tilde{\Delta} \sqrt{\beta_0} \beta_1^{3/2}}\,\,,\,\,A_{1\varphi}=\frac{\Delta}{\tilde{\Delta}}\,\frac{3m \left(\beta_1^2-1\right)}{2 \sqrt{2} \beta_1}\,\cos\theta \,.
\end{align}

\subsubsection{Extremal limits}
We now apply to the above solution the limit-prescriptions as given in the first part of  Sect.\ref{limitsec}.
It is useful to rewrite, in the $m\rightarrow 0$ limit, the $\eta_\ell$ in terms of harmonic functions:
\begin{equation}
H^0=1+\sqrt{2}\,\frac{|p^0|}{r}= \frac{\eta_0}r \,\,;\,\,\,H_1=1+\sqrt{2}\,\frac{|q_1|}{3\,r}= \frac{\eta_1}r\,,
\end{equation}
In the same limit we also have $\tilde{\Delta}\rightarrow r^2$.

\paragraph{non-BPS under-rotating limit\\}
This limit is obtained by redefining $\beta_0 = m\, \alpha_0$, $\beta_1 =  \alpha_1/m$ and $\alpha = m \Omega$ and then taking the limit $m\to 0$ in the solution. The charges at infinity become:
\begin{eqnarray}\label{chargesn}
&p^0= \frac{-1}{2\sqrt 2\alpha_0}\,,\quad q_0=0\,,\quad p^1=0\,,\quad
 q_1=-\frac{3\,\alpha_1}{2\sqrt 2}\,,&\nonumber\\
&M_{ADM}=-\frac{p^0+q_1}{2\,\sqrt 2}\,,\quad   M_\varphi=\Omega \frac{\sqrt{p^0\,q_1^3}}{3 \sqrt 3}=\frac{\Omega}{2}\,\sqrt{|I_4(p,q)|}\,,&\nonumber\\
  &\Sigma_1=0\,,\quad
  \Sigma_2 =\,\frac{q_1-3\,p^0}{2\,\sqrt 6}\,.&
\end{eqnarray}
and the four dimensional fields can be written  as
\begin{eqnarray}
% \nonumber to remove numbering (before each equation)
 e^{-4U} &=& H^0\,H_1^3-4 M_\varphi^2 \frac{\cos^2\theta}{r^4}\,, \quad
\omega = 2\, M_\varphi \,\frac{\sin^2\theta}r\,d\varphi\,,\nonumber\\
 T&=&\frac{1}{H^0\,H_1}\left(2\,
\frac{M_\varphi\,\cos\theta }{r^2}-{\rm i}\,e^{-2U}\right)\,,\nonumber\\
A^0&=&-\,\sqrt 2 \,\frac{e^{4 U}}{r^2}\, M_\varphi\, \cos\theta\, H^0 \,(dt + \omega)- p^0\cos\theta d\varphi \\
 A^1 &=& -\frac{e^{4 U}}{r}\,\left(\frac{q_1}3\,H^0\,H_1^2+\frac{2\sqrt 2}{r^3} M_\varphi^2\,\cos^2\theta\right)\,(dt + \omega)- \frac {\sqrt 2 M_\varphi \sin^2\theta}{r} d\varphi \\
  A_0 &=& \,\frac{e^{4 U}}r\, \left(p^0 \,H_1^3+\,\frac{2\sqrt 2}{r^3}M_\varphi^2\,\cos^2\theta\right)(dt + \omega)+\frac {\sqrt 2 M_\varphi \sin^2\theta}{r} d\varphi \\
A_1 &=& -\,3\,\sqrt 2 \frac{e^{4 U}}{r^2}\, H_1\, M_\varphi\, \cos\theta (dt + \omega)- q_1\cos\theta d\varphi\,.
\end{eqnarray}
\paragraph{Rotating-BPS limit}

This limit is obtained by redefining $\beta_0 = m\, \alpha_0$, $\beta_1 = m \, \alpha_1$  and then taking the limit $m\to 0$ in the solution keeping $\alpha$ fixed.

In this limit the charges at infinity become:
\begin{eqnarray}
M_{ADM}&=& \frac{1}{8} \left(\frac{3}{\alpha _1}+\frac{1}{\alpha _0}\right)=\frac{q_1-p^0}{2\sqrt{2}}\,,\quad p^0 = -\frac{1}{2 \sqrt{2} \alpha _0}\,,\quad q_1 = \frac{3}{2 \sqrt{2} \alpha _1}\,\,,\,\,\Sigma_1=0\,,\nonumber\\
\Sigma_2 &=& \frac{\sqrt{3} \left(\alpha _1-\alpha _0\right)}{8 \alpha _0 \alpha _1}=-\frac{q_1+3p^0}{2\sqrt{6}}\,,\quad M_\varphi=\frac{\alpha\,\left(\alpha _0+3 \alpha _1\right)}{8 \sqrt{\alpha _0} \alpha _1^{3/2}}=\frac{\alpha}{6\sqrt{6}}\sqrt{-\frac{q_1}{p^0}}\,(q_1-9 p^0)\,.
\end{eqnarray}
and the four dimensional fields can be written as
\begin{align}
e^{-4U}&=\frac{\tilde{\rho}^4}{\tilde{\Delta}^2}\,,\nonumber\\
\tilde{\rho}^4&=r^4\,\left([H^0 H_1+\frac{\alpha^2}{r^2}\,\cos^2\theta ][ H_1^2+\frac{\alpha^2}{r^2}\,\cos^2\theta ]+\frac{\alpha^2\,\cos^2\theta }{54\, r^4}\,(3p^0+q_1)^2\frac{q_1}{p^0}\right)\,,\nonumber\\
\tilde{\Delta}^2&=r^2+\alpha^2\,\cos^2\theta \,\,;\,\,\,
\omega=\frac{ \sin^2\theta}{\tilde{\Delta}}\,(2\,M_\varphi r+{\alpha}\,\sqrt{I_4(p,q)})\,.\nonumber\\
T&=\frac{1}{r^2 H_1^2+{\alpha^2}\,\cos^2\theta }\,\left(\frac{\alpha \cos\theta (3p^0+q_1)}{3\sqrt{6} }\sqrt{-\frac{q_1}{p^0}}-{\rm i}\,e^{-2U}\,\tilde{\Delta}\right)\,,\nonumber\\
A^0&=\frac{r^2\alpha\cos\theta }{6\sqrt{3}\,\tilde{\rho}^4}\,(9p^0+q_1)\sqrt{-\frac{q_1}{p^0}}\left(H^0H_1+\frac{\alpha^2}{r^2}\cos^2\theta -\frac{\sqrt{2}\,q_1\,H^0}{3r}\frac{(3p^0+q_1)}{(9p^0+q_1)}\right)(dt+\omega)-\nonumber\\
&-\frac{\Delta }{\tilde{\Delta}}\,p^0\cos\theta \,d\varphi\,,\nonumber\\
A^1&=-\frac{q_1\,r^3}{3\tilde{\rho}^4}\,\left(H_1[H_1 H^0+\frac{\alpha^2}{r^2}\cos^2\theta ]-\frac{\alpha^2 [9(p^0)^2-q_1^2]}{18\sqrt{2}\,p^0 r^3}\cos^2\theta  \right)\,(dt+\omega)-\nonumber\\
&-\frac{\alpha\,\sin^2\theta}{6\sqrt{6}\tilde{\Delta}}[4p^0\,q_1+\sqrt{2}r(3p^0-q_1)]\,\sqrt{-\frac{q_1}{p^0}}\,d\varphi\,,\nonumber\\
A_0&=\frac{p^0\,r^3}{\tilde{\rho}^4}\,\left(H_1[H_1^2+\frac{\alpha^2}{r^2}\cos^2\theta ]-\frac{\alpha^2\,q_1\, [9(p^0)^2-q_1^2]}{54\sqrt{2}\,(p^0)^2 r^3}\cos^2\theta  \right)\,(dt+\omega)-\nonumber\\
&-\frac{\alpha\,\sin^2\theta}{6\sqrt{6}\tilde{\Delta}}[4p^0\,q_1+\sqrt{2}\,r\,(9p^0+q_1)]\,\sqrt{-\frac{q_1}{p^0}}\,
d\varphi\,,\nonumber\\
A_1&=-\frac{r^2\alpha\cos\theta }{2\sqrt{3}\,\tilde{\rho}^4}\,(3p^0-q_1)\sqrt{-\frac{q_1}{p^0}}\left(H_1^2+\frac{\alpha^2}{r^2}\cos^2\theta +\frac{\sqrt{2}\,q_1\,H_1}{3r}\frac{(3p^0+q_1)}{(3p^0-q_1)}\right)(dt+\omega)-\nonumber\\
&-\frac{\Delta }{\tilde{\Delta}}\,q_1\cos\theta \,d\varphi\,.
\end{align}
Setting $\alpha\rightarrow 0$ we recover the known (regular) static BPS solution \cite{Behrndt:1997ny}.
\section{Aknowledgements}
M.T. wishes to thank Armen Yeranyan and Bert Vercnocke for interesting discussions.
This work was partially supported by
the Italian MIUR-PRIN contract 2009KHZKRX-007 "Symmetries of the Universe and of
the Fundamental Interactions".

\appendix

\section{Coset Geometry of the $D=3$ Sigma-Model}\label{A1}
 The sigma-model scalar fields $(\phi^I)\equiv \{U,a,\phi^s,\mathcal{Z}^M\}$ span a homogeneous-symmetric, pseudo-Riemannian  scalar manifold $\mathcal{M}_{scal}$ of the form
\begin{equation}
\mathcal{M}_{scal}=\frac{G_{(3)}}{H^*}\,.\label{scalma}
\end{equation}
The isometry group $G_{(3)}$ of the target space is the global symmetry group of $\mathcal{L}_{(3)}$
and $H^*$ is a suitable non-compact semisimple maximal subgroup of it.
We shall use for this manifold the solvable Lie algebra parametrization by identifying the scalar
fields $\phi^I$ with parameters of a suitable solvable Lie algebra \cite{Chemissany:2010zp}.
 Indeed  the scalars
$\phi^I$ define a \emph{local} solvable parametrization, i.e. the
corresponding patch, to be dubbed \emph{physical patch} ${\Scr U}$, is isometric to a solvable Lie group
generated by a solvable Lie algebra $Solv$:
\begin{equation}
\mathcal{M}_{scal}\supset {\Scr U}\equiv e^{Solv}\,,
\end{equation}
$Solv$ is defined by the Iwasawa decomposition of the Lie algebra
$\mathfrak{g}$ of $G_{(3)}$ with respect to its maximal compact
subalgebra $\mathfrak{H}$. The solvable parametrization $\phi^I$
can be defined by the following exponential map:
\begin{equation}
\mathbb{L}(\phi^I)=\exp(-a T_\bullet)\,\exp(\sqrt{2}
\mathcal{Z}^M\,T_M)\,\exp(\phi^r\,T_r)\,\exp(2\,U
H_0)\,,\label{cosetr3}
\end{equation}
where the  generators $H_0,\,T_\bullet,\,T_r,\,T_M$ satisfy the
following commutation relations:
\begin{align}
[H_0,\,T_M]&=\frac{1}{2}\,T_M\,\,;\,\,\,[H_0,\,T_\bullet]=T_\bullet\,\,;\,\,\,[T_M\,T_N]=\mathbb{C}_{MN}\,T_\bullet\,,\nonumber\\
[H_0,T_r]&=[T_\bullet,T_r]=0\,\,;\,\,\,[T_r,T_M]=T_r{}^N{}_M\,T_N\,\,;\,\,\,[T_r,T_s]=-
T_{rs}{}^{s'} T_{s'}\,,\label{relc1}
\end{align}
$T_r{}^N{}_M$ representing the symplectic ${\bf R}$ representation of $T_r$ on
contravariant symplectic vectors $d\mathcal{Z}^M$.\par
In all  $\mathcal{N}=2$ models with just vector multiplets $n_v=n_s/2+1$ and thus the dimension of the scalar manifold in $D=3$ is $4\,n_v$. This manifold is a pseudo-quaternionic K\"ahler space.
\par
 The coset geometry is defined by  the
 involutive automorphism $\sigma$ on the algebra
$\mathfrak{g}$ of $G_{(3)}$ which leaves the algebra $\mathfrak{H}^*$
generating $H^*$ invariant. All the formulas related to the group $G_{(3)}$ and its generators are referred to a matrix representation of $G_{(3)}$ (we shall in particular use the fundamental one). The involution $\sigma$ in the chosen representation has the general action: $\sigma(M)=-\eta M^\dagger\eta$, $\eta$
being an $H^*$-invariant metric ($\eta=\eta^\dagger,\,\,\eta^2={\bf 1}$), and induces the (pseudo)-Cartan
decomposition of $\mathfrak{g}$ of the form:
\begin{equation}
\mathfrak{g}=\mathfrak{H}^*\oplus \mathfrak{K}^*\,,\label{pseudoC}
\end{equation}
where $\sigma(\mathfrak{K}^*)=-\mathfrak{K}^*$, and the following
relations hold
\begin{equation}[\mathfrak{H}^*,\mathfrak{H}^*]\subset\mathfrak{H}^*,
\quad [\mathfrak{H}^*,\mathfrak{K}^*]\subset \mathfrak{K}^*,\quad
[\mathfrak{K}^*,\mathfrak{K}^*] \subset
\mathfrak{H}^*.\label{HKrels}\end{equation}
 We see that $H^*$ has a
linear adjoint action  in the space $\mathfrak{K}^*$ which is thus the
carrier of an $H^*$-representation. A general feature of
$\mathcal{N}=2$ symmetric models is that the isotropy group has the form $H^*={\rm SL}(2,\mathbb{R})\times
G'_4$ and its adjoint action on $\mathfrak{K}^*$ realizes the
representation ${\bf (2,R)}$.\par
The decomposition (\ref{pseudoC}) has to be contrasted with the ordinary Cartan decomposition of $\mathfrak{g}$
\begin{equation}
\mathfrak{g}=\mathfrak{H}\oplus \mathfrak{K}\,,\label{Cartan}
\end{equation}
into its maximal compact subalgebra $\mathfrak{H}$ generating $H$ and its orthogonal non-compact complement $\mathfrak{K}$. This decomposition is effected through the Cartan involution $\tau$ of which $\mathfrak{H}$  and $\mathfrak{K}$ represent the eigenspaces with eigenvalues $+1$ and $-1$ respectively. In the matrix representation in which we shall work, the action of $\tau$ can be implemented as: $\tau(X)=-X^\dagger$. We shall also use the $H^*$-invariant symetric matrix $\mathcal{M}(\Phi^I)=\mathbb{L}(\Phi^I)\eta\,\mathbb{L}(\Phi^I)^\dagger$.
\par Next we construct
the left invariant one-form and the vielbein
$P^{\mathcal{A}}=P_I{}^{\mathcal{A}} d\phi^I$:
\begin{equation}
\mathbb{L}^{-1}d\mathbb{L}=P^{\mathcal{A}}\,T_{\mathcal{A}}=P+W\,\,;\,\,\,{\mathcal{A}}=1,\dots,
{\rm dim}(\mathcal{M}_{scal})\,.\label{li1f}
\end{equation}
where $P=P^{\mathcal{A}} \mathbb{K}_{\mathcal{A}}$ and $W$ are the vielbein and connection matrices,  $\{\mathbb{K}_\mathcal{A}\}$ being a basis of
$\mathfrak{K}^*$ defined as follows:
\begin{equation}
\mathbb{K}_\mathcal{A}=\frac{1}{2}\,(T_\mathcal{A}+\eta\,T_\mathcal{A}^\dagger\,\eta)\,,
\end{equation}
where $T_\mathcal{A}=T_I$ are the solvable generators defined above.\par
Following the prescription of \cite{Chemissany:2010zp}, the
normalization of the $H^*$-invariant metric on the tangent space of
$G_{(3)}/H^*$ is chosen as follows
\begin{equation}
g_{{\mathcal{A}}{\mathcal{B}}}=k\, {\rm
Tr}[\mathbb{K}_{\mathcal{A}}\mathbb{K}_{\mathcal{B}}]\,,
\end{equation}
where $k=1/(2 {\rm Tr}(H_0^2))$ is a representation-dependent constant. \par The metric of the $D=3$
sigma-model has the familiar form:
\begin{align}
dS^2&=k\, {\rm
Tr}[P^2]=P^{\mathcal{A}} P^{\mathcal{B}}
g_{{\mathcal{A}}{\mathcal{B}}}=2 dU^2+ g_{rs} d\phi^r
d\psi^s+\frac{e^{-4U}}{2} \omega^2+e^{-2U}
d\mathcal{Z}^T\mathcal{M}_4(\phi^r)d\mathcal{Z}\,,\\
\omega&= da+ \mathcal{Z}^T\mathbb{C}d\mathcal{Z}\,,
\end{align}
\subsection{The STU model.}
The STU model is an $\mathcal{N}=2$ supergravity coupled to three vector multiplets ($n_s=6,\,n_v=4$) and with:
\begin{equation}
\mathcal{M}^{(D=4)}=\left(\frac{{\rm SL}(2,\mathbb{R})}{{\rm SO}(2)}\right)^3\,.
\end{equation}
This manifold is a complex spacial K\"ahler space spanned by three complex scalar fields $z^a=\{S,T,U\}$.
The $D=4$ scalar metric for the STU model reads
\begin{equation}
dS^2_4=g_{rs}\,d\phi^sd\phi^r=2\,g_{a\bar{b}}dz^ad\bar{z}^{\bar{b}}=-2\,\sum_{a=1}^3\frac{dz^{a}
d\bar{z}^{\bar{a}}}{(z^a-\bar{z}^{\bar{a}})^2}\,.
\end{equation}
Upon timelike reduction to $D=3$ the scalar manifold has the form (\ref{scalma}) with $G_{(3)}={\rm SO}(4,4)$ and $H^*={\rm SO}(2,2)^2$.
\subsection{The $T^3$ model.} \label{t3}
The $T^3$ model is an $\mathcal{N}=2$ supergravity coupled to a single  vector multiplet ($n_s=2,\,n_v=2$) and with scalar manifold
\begin{equation}
\mathcal{M}^{(D=4)}=\frac{{\rm SL}(2,\mathbb{R})}{{\rm SO}(2)}\,,
\end{equation}
spanned by a single complex scalar field $T=y-i\,e^{\varphi}$. It originates from the pure $D=5$ supergravity with the same amout of supersymmetries.
The $D=4$ scalar metric for the $T^3$-model is
  \begin{equation}
dS^2_4=g_{rs}\,d\phi^sd\phi^r=2\,g_{a\bar{b}}dz^ad\bar{z}^{\bar{b}}=-6\,\frac{dT
d\bar{T}}{(T-\bar{T})^2}\,.
\end{equation}
The scalar manifold in $D=3$ has the form (\ref{scalma}) with $G_{(3)}={\rm G}_{2(2)}$ and $H^*={\rm SL}(2,\mathbb{R})^2$.
\section{The $\mathfrak{g}_{2(2)}$ Lie Algebra in Terms of Chevalley Triples}\label{A2}
The complex Lie algebra $\mathfrak{g}_2(\mathbb{C})$ has rank two and it is defined by the $2\times 2$ Cartan matrix encoded in the following Dynkin diagram:
\begin{center}
\begin{picture}(110,30)
\put (-60,20){$\mathfrak{g}_2$}
\put (10,23){\circle {10}}
\put (15,25){\line (1,0){20}}
\put (15,23){\line (1,0){20}}
\put (20.5,19){{\LARGE$>$}}
\put (15,20.5){\line (1,0){20}}
\put (40,23){\circle {10}}
\put (65,21){$=\quad\quad\left (\begin{array}{cc}
  2 & -3\\
  -1 & 2
\end{array} \right)$}
\end{picture}
\end{center}
The $\mathfrak{g}_2$ root system $\Delta$ consists of the following six positive roots plus their negatives:
\begin{equation}
\label{g2rootsystem}
\begin{array}{rclcrcl}
\alpha_1&=&(1,0)&;&\alpha_2&=&\frac{\sqrt{3}}{2}\,(-\sqrt{3},1)\\
\alpha_3 \, =\, \alpha_1+\alpha_2&=&\frac{1}{2}\,(-1,\sqrt{3}) &;&
\alpha_4 \, = \, 2\,\alpha_1+\alpha_2 &=&
\frac{1}{2}\,(1,\sqrt{3}) \\
\alpha_5 \, = \, 3\,\alpha_1+\alpha_2&=&\frac{\sqrt{3}}{2}\,(\sqrt{3},1)&;&\alpha_6 \, = \,  3\,\alpha_1+2\,\alpha_2 &=&
(0,\sqrt{3})\
\end{array}
\end{equation}
Let  $\{H_1,\,H_2\}$ be the  Cartan generators along the two ortho-normal
directions and denote by $E^{\pm \alpha}$ the shift generators corresponding to the positive root $\alpha$ and its negative, normalized according to the standard Cartan--Weyl conventions:
\begin{eqnarray}
[E^\alpha,E^{-\alpha}]&=&\alpha^i\,H_i\,\,,\,\,\,[H_i,E^\alpha]=\alpha^i\,E^\alpha\,.
\end{eqnarray}
It is convenient to write the $\mathfrak{g}_{2(2)}$ generators and their commutation relations in terms of triples of Chevalley generators.
\par
Since the algebra has rank two there are two fundamental triples of Chevalley generators:
\begin{equation}\label{fundatriplets}
    \left(h_1,e_1,f_1\right ) \quad ; \quad \left(h_2,e_2,f_2\right )
\end{equation}
with the following commutation relations:
\begin{equation}\label{fundacommu}
    \begin{array}{ccccccc}
       \left[h_2,e_2\right]=2e_2 & \null &\left[h_1,e_2\right]=-3e_2& \null & \left[h_2,f_2\right]=-2f_2 & \null & \left[h_1,f_2\right]=3f_2  \\
       \left[h_2,e_1\right]=-e_1 & \null &\left[h_1,e_1\right]=
       2e_1& \null & \left[h_2,f_1\right]=f_1 & \null & \left[h_1,f_1\right]=-2f_1 \\
       \left[e_2,f_2\right]=h_2 & \null &\left[e_2,f_1\right]=0 & \null &
       \left[e_1,f_1\right]=h_1 & \null & \left[e_1,f_2\right]=0 \\
     \end{array}
\end{equation}
The remaining basis elements are defined as follows:
\begin{equation}\label{ancora}
    \begin{array}{ccccccc}
       e_3=\left[e_1,e_2\right]& \null & e_4=\frac{1}{2}\,\left[e_1,e_3\right] & \null &
       e_5=\frac{1}{3}\,\left[e_4,e_1\right] & \null & e_6=\left[e_2,e_5\right] \\
       f_3=\left[f_2,f_1\right]& \null & f_4=\frac{1}{2}\,\left[f_3,f_1\right] & \null &
       f_5=\frac{1}{3}\,\left[f_1,f_4\right] & \null & f_6=\left[f_5,f_2\right] \\
     \end{array}
\end{equation}
and satisfy the following Serre relations:
\begin{equation}\label{serrerele}
    \left [ e_2,e_3\right] \,=\,\left [ e_5,e_1\right] \,=\,\,\left [ f_2,f_3\right] \,=\,\left [ f_5,f_1\right] \,=\,0
\end{equation}
The Chevalley form of the commutation relation is obtained from the standard Cartan Weyl basis introducing the following identifications:
\begin{equation}\label{chevallabasa}
    \begin{array}{ccccccc}
       e_1 & = & \sqrt{2} E^{\alpha_1}& ; & e_2 & = &\sqrt{\frac{2}{3}} E^{\alpha_2} \\
       e_3 & = & \sqrt{2} E^{\alpha_3} & ; &  e_4 & =& \sqrt{2} E^{\alpha_4} \\
       e_5 & = & \sqrt{\frac{2}{3}} E^{\alpha_5} & ; & e_6 & = & \sqrt{\frac{2}{3}} E^{\alpha_6}\\
       f_1 & = & \sqrt{2} E^{-\alpha_1}& ; & f_2 & = &\sqrt{\frac{2}{3}} E^{-\alpha_2} \\
       f_3 & = & \sqrt{2} E^{-\alpha_3} & ; &  f_4 & = &\sqrt{2} E^{-\alpha_4} \\
       f_5 & = & \sqrt{\frac{2}{3}} E^{-\alpha_5} &  ; & f_6 & = & \sqrt{\frac{2}{3}} E^{-\alpha_6}
     \end{array}
\end{equation}
and\footnote{Note that we are using a slightly different notation with respect to \cite{Kim:2010bf}: Denoting by bold symbols the Chevalley generators used in that reference we have the following correspondence:
\begin{eqnarray}
e_1&=& {\bf e}_2\,;\,\,e_2={\bf e}_1\,;\,\,e_3=-{\bf e}_3\,;\,\,e_4={\bf e}_4/2\,;\,\,e_5={\bf e}_5/6\,;\,\,e_6={\bf e}_6/6\,;\,\, h_1={\bf h}_2\,;\,\, h_2={\bf h}_1\,.
\end{eqnarray} }
\begin{equation}\label{cartanucci}
    h_1 \, = \, 2 \alpha_1\cdot H \quad ; \quad h_2 \, = \, \frac{2}{3} \, \alpha_2\cdot H
\end{equation}
The solvable generators $T_\mathcal{A}$ have the following expression in the Chevalley basis:
\begin{align}
H_0&=\frac{h_1}{2}+h_2\,\,;\,\,\,T_{r=1}=e_1\,\,;\,\,\,T_{r=2}=\frac{h_1}{2}\,\,;\,\,\,T_{0}=e_2\,\,;\,\,\,T_1=e_3\,\,;\,\,\,T^0=e_5\,\,;
\,\,\,T^1=\frac{e_4}{3}\,\,;\,\,\,T_\bullet=e_6\,,
\end{align}
The parametrization of the $D=3$ scalar manifold is defined by the coset prepresentative
\begin{equation}
\mathbb{L}(\phi^I)=\exp(-a T_\bullet)\,\exp(\sqrt{2}
\mathcal{Z}^M\,T_M)\,\exp(y\,T_{r=1})\,\exp(\varphi\,T_{r=2})\exp(2\,U
H_0)\,,\label{cosetr4}
\end{equation}
\subsection{Normal Form of the Electric-Magnetic Charge Representation}\label{A22}
The normal-form spaces $\mathbb{J}^{(N)}$ and $\mathbb{K}^{(N)}$ are $p=2$-dimensional, $p$ being the rank of the coset $H^\star/H_c={\rm SL}(2)^2/{\rm SO}(2)^2$. The two parameters of the normal form of the representation ${\bf R}$ of $H_c$ can either be related to the set of  $D0-D4$ charges $q_0,\,p^1$ or to the  $D2-D6$ charges
$q_1,\,p^0$. According to our discussion in Sect. \ref{section1}, the generators $\mathcal{J}_\ell$ and $\mathcal{K}_\ell$ of  $\mathbb{J}^{(N)}$ and $\mathbb{K}^{(N)}$, respectively, together with $\mathcal{H}_\ell\equiv [\mathcal{K}_\ell,\,\mathcal{J}_\ell]$, generate a characteristic  ${\rm
SL}(2,\mathbb{R})^2$ subgroup of ${\rm G}_{2(2)}$.
 These generators  are constructed out of
the nilpotent matrices $T_M$ corresponding to these charges. In the
case of the normal form $q_0,\,p^1$ we write  ${\rm
SL}(2,\mathbb{R})^2={\rm SL}(2,\mathbb{R})_{q_0}\times {\rm
SL}(2,\mathbb{R})_{p^1}$, so that the two factor groups are
generated by the following algebras:
\begin{eqnarray}
\mathfrak{sl}(2,\mathbb{R})_{q_0}&\equiv&\{\mathcal{J}_{q_0},\,\mathcal{K}_{q_0},\mathcal{H}_{q_0}\}\,\,:\,\,\begin{cases}
{\cal J}_{q_0}=\frac{T_0+T_0^T}{2}=\frac{e_2+f_2}{2}\cr \mathcal{K}_{q_0}=\frac{T_0-T_0^T}{2}=\mathbb{K}_0=\frac{e_2-f_2}{2}\cr \mathcal{H}_{q_0}=\frac{h_2}{2}\end{cases} \nonumber\\
\mathfrak{sl}(2,\mathbb{R})_{p^1}&\equiv&\{{\cal J}_{p^1},\,\mathcal{K}_{p^1},\mathcal{H}_{p^1}\}\,\,:\,\,\begin{cases}
{\cal J}_{p^1}=3\,\frac{T^1+T^{1\,T}}{2}=\frac{e_4+f_4}{2}\cr
\mathcal{K}_{p^1}=3\,\frac{T^1-T^{1\,T}}{2}=3\,\mathbb{K}^1=\frac{e_4-f_4}{2}\cr
\mathcal{H}_{p^1}=h_1+\frac{3 h_2}{2}\end{cases}\label{q0p1}
\end{eqnarray}
where the normalizations are chosen so that, defining for each algebra the nilpotent generators $N^\pm_{Q_0},\,N^\pm_{P^1}$ as:
\begin{eqnarray}
N^\pm_{q_0}&\equiv & \mathcal{H}_{q_0}\mp
\mathcal{K}_{q_0}\,\,;\,\,\,N^\pm_{p^1}\equiv \mathcal{H}_{p^1}\mp \mathcal{K}_{p^1}\,,
\end{eqnarray}
the following commutation relations hold:
\begin{align}
[K,\,J]&=\mathcal{H}\,,\,\,\,[\mathcal{H},K]=J\,,\,\,\,[\mathcal{H},J]=K\,,\,\,\,
[J,N^\pm]=\pm N^\pm\,,\,\,\,[N^+,\,N^-]=2\,J\,,
\end{align}
where we have suppressed the charge subscripts and it is easily
verified that generators with different subscripts, thus pertaining
to different $\mathfrak{sl}(2)$ algebras, commute. The matrices
$\mathcal{J}_{p^1},\,\mathcal{J}_{q_0}$ are non-compact and generate Harrison
transformations which, acting on a neutral solution (e.g.
Schwarzshild or Kerr) switch on the charges $q_0,\,p^1$.
\par We can alternatively choose the normal form $\{p^0,\,q_1\}$. In
this case we write ${\rm SL}(2,\mathbb{R})^2={\rm
SL}(2,\mathbb{R})_{p^0}\times {\rm SL}(2,\mathbb{R})_{q_1}$, where:
\begin{eqnarray}
\mathfrak{sl}(2,\mathbb{R})_{p^0}&\equiv&\{\mathcal{J}_{p^0},\,\mathcal{K}_{p^0},\mathcal{H}_{p^0}\}\,\,:\,\,\begin{cases}
\mathcal{J}_{p^0}=\frac{T^0+T^{0T}}{2}=\frac{e_5+f_5}{2}\cr
\mathcal{K}_{p^0}=\frac{T^0-T^{0T}}{2}=\mathbb{K}^0=\frac{e_5-f_5}{2}\cr
\mathcal{H}_{p^0}=\frac{h_1+h_2}{2}
\end{cases} \nonumber\\
\mathfrak{sl}(2,\mathbb{R})_{q_1}&\equiv&\{\mathcal{J}_{q_1},\,\mathcal{K}_{q_1},\mathcal{H}_{q_1}\}\,\,:\,\,\begin{cases}
\mathcal{J}_{q_1}=\frac{T_1+T_1^{T}}{2}=\frac{e_3+f_3}{2}\cr
\mathcal{K}_{q_1}=\frac{T_1-T_1^{T}}{2}=\mathbb{K}_1=\frac{e_3-f_3}{2}\cr
\mathcal{H}_{q_1}=\frac{h_1+3 h_2}{2}\end{cases}
\end{eqnarray}
Similarly to the previous case, the action of Harrison
transformations generated by $\{\mathcal{J}'_\ell\}=\{\mathcal{J}_{p^0},\,\mathcal{J}_{q_1}\}$ on a neutral
solution generates the charges $p^0,\,q_1$.\par Let us choose for the
moment the $\{q_0,\,p^1\}$ normal form and use
$\{\mathcal{J}_\ell\}=\{\mathcal{J}_{q_0},\,\mathcal{J}_{p^1}\}$ as the generators of a
Cartan subalgebra $\mathcal{C}$ of $\mathfrak{g}_{2(2)}$. With
respect to $\mathcal{C}$, the space $\mathfrak{K}^*$ is spanned by a
basis of positive-root shift generators and their negatives. In
particular, denoting by  $\mathcal{N}_{a,b}$ a  nilpotent generator
in $\mathfrak{K}^*$ with grading $a$ with respect to $\mathcal{J}_1$
and $b$ relative to $\mathcal{J}_2$, \begin{equation}
[\mathcal{J}_1,\,\mathcal{N}_{(a,b)}]=a\,\mathcal{N}_{(a,b)}\,,\,\,\,[\mathcal{J}_2,\,\mathcal{N}_{(a,b)}]=b\,\mathcal{N}_{(a,b)}\,,
\end{equation}
the coset space $\mathfrak{K}^*$ is spanned by the following
generators:
\begin{equation}
\mathfrak{K}^*=\mbox{Span}(\mathcal{N}_{\left(\frac{1}{2},\,\frac{3}{2}\right)},\,\mathcal{N}_{\left(\frac{1}{2},\,-\frac{1}{2}\right)},\,
\mathcal{N}_{(1,\,0)},\,\mathcal{N}_{(0,\,1)},\,\mathcal{N}_{\left(-\frac{1}{2},\,-\frac{3}{2}\right)},\,\mathcal{N}_{\left(-
\frac{1}{2},\,\frac{1}{2}\right)},\,
\mathcal{N}_{(-1,\,0)},\,\mathcal{N}_{(0,\,-1)})\,,\label{Nbasis2}
\end{equation}
where
\begin{align}
\mathcal{N}_{(\pm
1,\,0)}&=N_{q_0}^{\pm}\,\,,\,\,\,\mathcal{N}_{(0,\,\pm
1)}=N_{p^0}^{\pm}\,,\nonumber\\
\mathcal{N}_{\left(\frac{1}{2},\,\frac{3}{2}\right)}&=\frac{1}{4}\left(e_1-e_3+e_5+e_6+f_1+f_3-f_5+f_6)\right)\,,\nonumber\\
\mathcal{N}_{\left(\frac{1}{2},\,-\frac{1}{2}\right)}&=\frac{1}{4}\left(-e_1+e_3+3e_5+3e_6-f_1-f_3-3f_5+3f_6)\right)\,.
\end{align}
This basis is convenient if we consider the Harrison transformation
$$\mathcal{O}_{(q_0,\,p^1)}\equiv
e^{\log(\beta_\ell)\,\mathcal{J}_\ell}\,\,\,,\,\,\,\,\,\beta_\ell>0$$
 which, on a neutral solution switches on the charges $q_0,\,p^1$.
Indeed each generator in the above basis  transform under this
Harrison boost by a scale factor:
\begin{equation}
\mathcal{O}_{(q_0,\,p^1)}^{-1}\,\mathcal{N}_{(a,b)}\,\mathcal{O}_{(q_0,\,p^1)}=\frac{1}{\beta_0^a\beta_1^b}\,
\mathcal{N}_{(a,b)}\,.\label{abaction}
\end{equation}
\section{The Kerr-Newmann-Taub-NUT Solution}\label{Afin}
The Kerr-Newmann-Taub-NUT solution has the general form \cite{stefani}:
%For stationary, axisymmetric, asymptotically flat solutions admitting the two
%Killing vectors $\partial_t$ and $\partial_\varphi$, the most general case of complex
%scalar fields $u,v$ corresponds to a Kerr--Newman solution with NUT-charge, whose metric reads
\begin{equation}
ds^2= \frac{\tilde\Delta}{\rho^2}(dt+\omega)^2-\frac{\rho^2}{\tilde\Delta}\left(\frac{\tilde\Delta}{\Delta} dr^2 + \tilde\Delta d\theta^2 +\Delta\sin^2\theta d\varphi^2\right)
\end{equation}
where
\begin{eqnarray}
\Delta &=& (r-m)^2 - c^2\,,\label{delta}\\
c^2 &=& m^2+\ell^2 -\frac 12 (q^2+p^2) -\alpha^2\,,\label{c2}\\
\tilde\Delta&=& \Delta -\alpha^2 \sin^2\theta\,,\label{deltatil}\\
\rho^2 &=& r^2+ \left(\alpha\cos\theta +\ell\right)^2\,,\label{rho}\\
\omega &=& \left(\alpha\sin^2\theta \frac{\rho^2-\tilde\Delta}{\tilde\Delta}+2\ell\cos\theta \right)d\varphi\,,\label{B}\\
A^0&=&\left[-q\,r+p\,(\ell+\alpha\,\cos\theta )\right]\frac{dt}{\rho^2}+\nonumber\\
&+&\{-p\left[(\alpha^2+r^2-\ell^2)\cos\theta +\alpha\ell\, \sin^2\theta \right]+q\,\left[\alpha\,r\,\sin^2\theta -2\,\ell\,r\,\cos\theta \right
]\}\,\frac{d\varphi}{\rho^2}\,,
\end{eqnarray}
 in terms of  the coordinates $(r,\theta)$,
 of the electric and magnetic charges $(q,p)$ and of
the ADM-mass and NUT charge $(m,\ell)$. The parameter $\alpha$, as before, is related to the angular momentum $M_\varphi$ of
the solution by $\alpha= M_\varphi/m$. Here the metric field $U(r,\theta)$ is given by $e^{2U}=
\frac{\tilde\Delta}{\rho^2}$.\par
The above Kerr-Newmann Taub-NUT solution, embedded in the $T^3$-model, is characterized by the following matrices $Q,\,Q_\psi$:
\begin{align}
Q&=\frac{1}{2}\,\left(-2\,\mathbb{K}_\bullet\,\ell + 4\,{H_0}\,m -
2\sqrt{2}\,\mathbb{K}_0\,q_0 + 2 \sqrt{2}\, \mathbb{K}^0\,p^0
-\frac{2\sqrt{2}}{3}\, \mathbb{K}_1\,q_1  +
  6\sqrt{2}\,\mathbb{K}^1\,p^1\right)\,,\label{KNNUT}\\
Q_\psi&=\frac{\alpha}{2}\left(4\,\mathbb{K}_\bullet\,m +
4\,{H_0}\,\ell + 2\sqrt{2}\,\mathbb{K}^0\,q_0 +
 2\sqrt{2}\, \mathbb{K}_0\,p^0  +2\sqrt{2}\, \mathbb{K}^1\,q_1  +
  2\sqrt{2}\,\mathbb{K}^1\,p^1\right)\,,\nonumber
\end{align}
where
\begin{equation}
M_{ADM}=m\,,\,\,p^0=\frac{p-q}{2\sqrt{2}}\,,\,\,p^1=\frac{p+q}{2\sqrt{2}}\,,\,\,q_0=\frac{p+q}{2\sqrt{2}}\,,\,\,q_1=3\frac{q-p}{2\sqrt{2}}\,,
\end{equation}
From Eq. (\ref{KNNUT}) we see that
the electric and magnetic charges of a solution can be read off the
components of $Q$ along the compact generators
$\mathbb{K}_M=(T_M-T_M^\dagger)/2$.\par


\begin{thebibliography}{99}
\bibitem{reviews} For reviews on black holes in superstring and supergravity theories see for example:
 J. M. Maldacena, ``Black-Holes in String Theory'', hep-th/9607235;
A.~W.~Peet, ``TASI lectures on black holes in string theory,''
  hep-th/0008241;
  %%CITATION = HEP-TH 0008241;%%
  B.~Pioline,
  ``Lectures on on black holes, topological strings and quantum attractors,''
  Class.\ Quant.\ Grav.\  {\bf 23} (2006) S981
  [arXiv:hep-th/0607227];
  %%CITATION = HEP-TH 0607227;%%
A.~Dabholkar,
  ``Black Hole Entropy And Attractors,''
  Class.\ Quant.\ Grav.\  {\bf 23} (2006) S957.
  %%CITATION = CQGRD,23,S957;%%
 L.~Andrianopoli, R.~D'Auria, S.~Ferrara and M.~Trigiante,
``Extremal black holes in supergravity,''
  Lect.\ Notes Phys.\  {\bf 737} (2008) 661
  [arXiv:hep-th/0611345];
  %%CITATION = LNPHA,737,661;%%
  \bibitem{attractors}
S.~Ferrara, R.~Kallosh,
  ``Supersymmetry and attractors'',
  Phys.\ Rev.\  {\bf D54}, 1514-1524 (1996).
  [hep-th/9602136];\\
  S.~Ferrara, R.~Kallosh, A.~Strominger,
 ``N=2 extremal black holes''
  Phys.\ Rev.\  {\bf D52 } (1995)  5412-5416.
[hep-th/9508072];
  G.~W.~Gibbons, R.~Kallosh, B.~Kol,
 ``Moduli, scalar charges, and the first law of black hole thermodynamics'',
  Phys.\ Rev.\ Lett.\  {\bf 77 } (1996)  4992-4995;
  K.~Goldstein, N.~Iizuka, R.~P.~Jena and S.~P.~Trivedi,
  \emph{Non-supersymmetric attractors}
  Phys.\ Rev.\  D {\bf 72}, 124021 (2005);\\
P.~K.~Tripathy and S.~P.~Trivedi,
  \emph{Non-Supersymmetric Attractors in String Theory}
  JHEP {\bf 0603}, 022 (2006);\\
R.~Kallosh,
  \emph{New Attractors}
  JHEP {\bf 0512}, 022 (2005)
  [arXiv:hep-th/0510024];\\
   D.~Astefanesei, K.~Goldstein, R.~P.~Jena, A.~Sen and S.~P.~Trivedi,
 ``Rotating attractors,''
  JHEP {\bf 0610} (2006) 058.
  %%CITATION = HEP-TH/0606244;%%
\bibitem{Goldstein:2008fq}
  K.~Goldstein and S.~Katmadas,
``Almost BPS black holes,''
  JHEP {\bf 0905} (2009) 058
  [arXiv:0812.4183 [hep-th]].
  %%CITATION = ARXIV:0812.4183;%%
 \bibitem{Bena:2009ev}  I.~Bena, G.~Dall'Agata, S.~Giusto, C.~Ruef and N.~P.~Warner,
  ``Non-BPS Black Rings and Black Holes in Taub-NUT,''
  JHEP {\bf 0906} (2009) 015
  [arXiv:0902.4526 [hep-th]];I.~Bena, S.~Giusto, C.~Ruef and N.~P.~Warner,
``Multi-Center non-BPS Black Holes: the Solution,''
  JHEP {\bf 0911} (2009) 032
  [arXiv:0908.2121 [hep-th]].
  %%CITATION = ARXIV:0908.2121;%%
 \bibitem{Bossard:2011kz}
  G.~Bossard and C.~Ruef,
  ``Interacting non-BPS black holes,''
  Gen.\ Rel.\ Grav.\  {\bf 44} (2012) 21
  [arXiv:1106.5806 [hep-th]];
  %%CITATION = ARXIV:1106.5806;%%
  G.~Bossard,
  ``Octonionic black holes,''
  JHEP {\bf 1205} (2012) 113
  [arXiv:1203.0530 [hep-th]].
  %%CITATION = ARXIV:1203.0530;%%
 \bibitem{Ferrara:2012qm}
  S.~Ferrara, A.~Marrani, A.~Shcherbakov and A.~Yeranyan,
 ``Multi-Centered First Order Formalism,''
  arXiv:1211.3262 [hep-th].
  %%CITATION = ARXIV:1211.3262;%%
 \bibitem{Cvetic:1995kv}
  M.~Cvetic and D.~Youm,
``All the static spherically symmetric black holes of heterotic string on a six torus,''
  Nucl.\ Phys.\ B {\bf 472} (1996) 249
  [hep-th/9512127].
  %%CITATION = HEP-TH/9512127;%%
  \bibitem{Breitenlohner:1987dg}
  P.~Breitenlohner, D.~Maison and G.~W.~Gibbons,
  ``Four-Dimensional Black Holes from Kaluza-Klein Theories,''
  Commun.\ Math.\ Phys.\  {\bf 120} (1988) 295.
  %%CITATION = CMPHA,120,295;%%
  \bibitem{Cvetic:1996kv}
  M.~Cvetic and D.~Youm,
``Entropy of nonextreme charged rotating black holes in string theory,''
  Phys.\ Rev.\ D {\bf 54} (1996) 2612
  [hep-th/9603147].
  %%CITATION = HEP-TH/9603147;%%
 \bibitem{Rasheed:1995zv}
  D.~Rasheed,
 ``The Rotating dyonic black holes of Kaluza-Klein theory,''
  Nucl.\ Phys.\ B {\bf 454} (1995) 379
  [hep-th/9505038];
  %%CITATION = HEP-TH/9505038;%%
   F.~Larsen,
 ``Rotating Kaluza-Klein black holes,''
  Nucl.\ Phys.\ B {\bf 575} (2000) 211
  [hep-th/9909102].
  %%CITATION = HEP-TH/9909102;%%
  %69 citations counted in INSPIRE as of 06 Mar 2013
\bibitem{Andrianopoli:2012ee}  L.~Andrianopoli, R.~D'Auria, P.~Giaccone and M.~Trigiante,
  ``Rotating black holes, global symmetry and first order formalism,''
  JHEP {\bf 1212} (2012) 078
  [arXiv:1210.4047 [hep-th]].
  %%CITATION = ARXIV:1210.4047;%%
\bibitem{pioline}
M.~Gunaydin, A.~Neitzke, B.~Pioline  and A.~Waldron, ``BPS
black holes,
  quantum attractor flows and automorphic forms'', Phys. Rev. {\bf D73} (2006)
  084019;
%%CITATION = HEP-TH/0512296;%%.
  M.~Gunaydin, A.~Neitzke, B.~Pioline, A.~Waldron,
 ``Quantum Attractor Flows'',
  JHEP {\bf 0709 } (2007)  056;
\bibitem{Gaiotto:2007ag} D.~Gaiotto, W.~W. Li  and M.~Padi, ``Non-Supersymmetric
Attractor Flow in
  Symmetric Spaces'', JHEP {\bf 12} (2007) 093,
%%CITATION = ARXIV:0710.1638;%%.
\bibitem{nilorbits} The main reference on this issue is D. H. Collingwood and
W.M. McGovern, \emph{Nilpotent Orbits in Semisimple Lie Algebras},  Van Nostrand Reinhold 1993.
\bibitem{Bergshoeff:2008be}
  E. Bergshoeff, W. Chemissany, A.~Ploegh, M.~Trigiante and T.~Van Riet,
 ``Generating Geodesic Flows and Supergravity Solutions,''
  Nucl.\ Phys.\ {\bf B 812} (2009) 343
  [arXiv:0806.2310].
  %%CITATION = NUPHA,B812,343;%%
\bibitem{Bossard:2009at}
  G.~Bossard, H.~Nicolai, K.~S.~Stelle,
 ``Universal BPS structure of stationary supergravity solutions'',
  JHEP {\bf 0907 } (2009)  003.
 \bibitem{Kim:2010bf}
  S.~-S.~Kim, J.~Lindman Hornlund, J.~Palmkvist and A.~Virmani,
  ``Extremal Solutions of the S3 Model and Nilpotent Orbits of G2(2),''
  JHEP {\bf 1008} (2010) 072
  [arXiv:1004.5242 [hep-th]].
  %%CITATION = ARXIV:1004.5242;%%

  \bibitem{marioetal} W. Chemissany, J. Rosseel, M. Trigiante, T. Van Riet,
   ``The full integration of black hole solutions to symmetric supergravity theories',
 Nucl. Phys. {\bf B 830} (2010) 391 [arXiv:0903.2777].
  \bibitem{Fre:2011uy}
  P.~Fre, A.~S.~Sorin and M.~Trigiante,
 ``Integrability of Supergravity Black Holes and New Tensor Classifiers of Regular and Nilpotent Orbits,''
  JHEP {\bf 1204} (2012) 015
  [arXiv:1103.0848 [hep-th]].
  %%CITATION = ARXIV:1103.0848;%%
\bibitem{Fre:2011ns}
  P.~Fre, A.~S.~Sorin and M.~Trigiante,
``Black Hole Nilpotent Orbits and Tits Satake Universality Classes,''
  arXiv:1107.5986 [hep-th].
  %%CITATION = ARXIV:1107.5986;%%
\bibitem{Chemissany:2012nb}
  W.~Chemissany, P.~Giaccone, D.~Ruggeri and M.~Trigiante,
``Black hole solutions to the $F_4$-model and their orbits (I),''
  Nucl.\ Phys.\ B {\bf 863} (2012) 260
  [arXiv:1203.6338 [hep-th]].
  %%CITATION = ARXIV:1203.6338;%%
  %3 citations counted in INSPIRE as of 28 Feb 2013
  \bibitem{Fre:2012im}
  P.~Fre and A.~S.~Sorin,
``Extremal Multicenter Black Holes: Nilpotent Orbits and Tits Satake Universality Classes,''
  JHEP {\bf 1301} (2013) 003
  [arXiv:1205.1233 [hep-th]].
  %%CITATION = ARXIV:1205.1233;%%
    \bibitem{Bouchareb:2007ax}
  A.~Bouchareb, G.~Clement, C.~-M.~Chen, D.~V.~Gal'tsov, N.~G.~Scherbluk and T.~Wolf,
 ``G(2) generating technique for minimal D=5 supergravity and black rings,''
  Phys.\ Rev.\ D {\bf 76} (2007) 104032
   [Erratum-ibid.\ D {\bf 78} (2008) 029901]
  [arXiv:0708.2361 [hep-th]].
  %%CITATION = ARXIV:0708.2361;%%
  \bibitem{Bertini:2011ga}
  S.~Bertini, S.~L.~Cacciatori and D.~Klemm,
  ``Conformal structure of the Schwarzschild black hole,''
  Phys.\ Rev.\ D {\bf 85} (2012) 064018
  [arXiv:1106.0999 [hep-th]].
  %%CITATION = ARXIV:1106.0999;%%
\bibitem{Cvetic:2011dn}
  M.~Cvetic and F.~Larsen,
  ``Conformal Symmetry for Black Holes in Four Dimensions,''
  JHEP {\bf 1209} (2012) 076
  [arXiv:1112.4846 [hep-th]].
  %%CITATION = ARXIV:1112.4846;%%

 \bibitem{Virmani:2012kw}
  A.~Virmani,
 ``Subtracted Geometry From Harrison Transformations,''
  JHEP {\bf 1207} (2012) 086
  [arXiv:1203.5088 [hep-th]].
  %%CITATION = ARXIV:1203.5088;%%
  \bibitem{Chakraborty:2012fx}
  A.~Chakraborty and C.~Krishnan,
 ``Attraction, with Boundaries,''
  arXiv:1212.6919 [hep-th].
  %%CITATION = ARXIV:1212.6919;%%
\bibitem{Cvetic:2013cja}
  M.~Cvetic, M.~Guica and Z.~H.~Saleem,
  ``General black holes, untwisted,''
  arXiv:1302.7032 [hep-th].
  %%CITATION = ARXIV:1302.7032;%%
  \bibitem{Bellucci:2006xz}
    S.~Bellucci, S.~Ferrara, M.~Gunaydin, A.~Marrani,
 ``Charge orbits of symmetric special geometries and attractors'',
  Int.\ J.\ Mod.\ Phys.\  {\bf A21 } (2006)  5043-5098.
  [hep-th/0606209].
  \bibitem{Behrndt:1997ny}
  K.~Behrndt, D.~Lust and W.~A.~Sabra,
 ``Stationary solutions of N=2 supergravity,''
  Nucl.\ Phys.\ B {\bf 510} (1998) 264
  [hep-th/9705169].
  %%CITATION = HEP-TH/9705169;%%

  \bibitem{Ferrara:2010ug}
  S.~Ferrara, A.~Marrani, E.~Orazi, R.~Stora and A.~Yeranyan,
 ``Two-Center Black Holes Duality-Invariants for stu Model and its lower-rank Descendants'',
  J.\ Math.\ Phys.\  {\bf 52} (2011) 062302.
\bibitem{Chemissany:2010zp}
  W.~Chemissany, P.~Fre, J.~Rosseel, A.~S.~Sorin, M.~Trigiante and T.~Van Riet,
``Black holes in supergravity and integrability,''
  JHEP {\bf 1009} (2010) 080
  [arXiv:1007.3209 [hep-th]].
  %%CITATION = ARXIV:1007.3209;%%



  \bibitem{stefani} H. Stephani, D. Kramer, M.A H. MacCallum, C. Hoenselaers and E. Herlt, ``Exact solutions
of Einstein's field equations'', Cambridge University Press, Cambridge, U.K. (2003).
\end{thebibliography}
\end{document}